\documentclass[10pt,preprint]{aastex}

\usepackage{graphicx,epsfig,subfigure,rotating,natbib}
\usepackage{amssymb,amsmath}

\newcommand{\myemail}{cathryn.trott@curtin.edu.au}

\slugcomment{Accepted by the Astrophysical Journal}

\shorttitle{CHIPS: The Cosmological HI Power Spectrum estimator}
\shortauthors{Trott et al.}

\begin{document}


\title{CHIPS: The Cosmological HI Power Spectrum Estimator}


\def\Curtin{\altaffilmark{1}}
\def\Curtintxt{\altaffiltext{1}{International Centre for Radio Astronomy Research, Curtin University, Perth, WA 6845, Australia}}

\def\CAASTRO{\altaffilmark{2}}
\def\CAASTROtxt{\altaffiltext{2}{ARC Centre of Excellence for All-sky Astrophysics (CAASTRO)}}

\def\UMelbourne{\altaffilmark{3}}
\def\UMelbournetxt{\altaffiltext{3}{The University of Melbourne, School of Physics, Parkville, VIC 3010, Australia}}

\def\UW{\altaffilmark{5}}
\def\UWtxt{\altaffiltext{5}{University of Washington, Department of Physics, Seattle, WA 98195, USA}}

\def\SKASA{\altaffilmark{6}}
\def\SKASAtxt{\altaffiltext{6}{Square Kilometre Array South Africa (SKA SA), Park Road, Pinelands 7405, South Africa}}

\def\RU{\altaffilmark{7}}
\def\RUtxt{\altaffiltext{7}{Department of Physics and Electronics, Rhodes University, Grahamstown 6140, South Africa}}

\def\CfA{\altaffilmark{8}}
\def\CfAtxt{\altaffiltext{8}{Harvard-Smithsonian Center for Astrophysics, Cambridge, MA 02138, USA}}

\def\ASU{\altaffilmark{9}}
\def\ASUtxt{\altaffiltext{9}{Arizona State University, School of Earth and Space Exploration, Tempe, AZ 85287, USA}}

\def\ANU{\altaffilmark{10}}
\def\ANUtxt{\altaffiltext{10}{Australian National University, Research School of Astronomy and Astrophysics, Canberra, ACT 2611, Australia}}

\def\Haystack{\altaffilmark{11}}
\def\Haystacktxt{\altaffiltext{11}{MIT Haystack Observatory, Westford, MA 01886, USA}}

\def\MIT{\altaffilmark{12}}
\def\MITtxt{\altaffiltext{12}{MIT Kavli Institute for Astrophysics and Space Research, Cambridge, MA 02139, USA}}

\def\Victoria{\altaffilmark{13}}
\def\Victoriatxt{\altaffiltext{13}{Victoria University of Wellington, School of Chemical \& Physical Sciences, Wellington 6140, New Zealand}}

\def\UWisc{\altaffilmark{14}}
\def\UWisctxt{\altaffiltext{14}{University of Wisconsin--Milwaukee, Department of Physics, Milwaukee, WI 53201, USA}}

\def\USydney{\altaffilmark{15}}
\def\USydneytxt{\altaffiltext{15}{The University of Sydney, Sydney Institute for Astronomy, School of Physics, NSW 2006, Australia}}

\def\CASS{\altaffilmark{4}}
\def\CASStxt{\altaffiltext{4}{CSIRO Astronomy and Space Science (CASS), PO Box 76, Epping, NSW 1710, Australia}}

\def\Tata{\altaffilmark{16}}
\def\Tatatxt{\altaffiltext{16}{National Centre for Radio Astrophysics, Tata Institute for Fundamental Research, Pune 411007, India}}

\def\RRI{\altaffilmark{17}}
\def\RRItxt{\altaffiltext{17}{Raman Research Institute, Bangalore 560080, India}}

\def\NRAO{\altaffilmark{18}}
\def\NRAOtxt{\altaffiltext{18}{National Radio Astronomy Observatory, Charlottesville and Greenbank, USA}}

\def\UWA{\altaffilmark{19}}
\def\UWAtxt{\altaffiltext{19}{International Centre for Radio Astronomy Research, University of Western Australia, Crawley, WA 6009, Australia}}

\def\ASTRON{\altaffilmark{20}}
\def\ASTRONtxt{\altaffiltext{20}{Netherlands Institute for Radio Astronomy (ASTRON), PO Box 2, 7990 AA Dwingeloo, The Netherlands}}

\def\UWeSci{\altaffilmark{21}}
\def\UWeScitxt{\altaffiltext{21}{University of Washington, eScience Institute, Seattle, WA 98195, USA}}

\def\Brown{\altaffilmark{22}}
\def\Browntxt{\altaffiltext{22}{Brown University, Department of Physics, Providence, RI 02912, USA}}

\def\myemail{\altaffilmark{*}}
\def\myemailtxt{\altaffiltext{*}{e-mail: cathryn.trott@curtin.edu.au}}

\author{
C.~M.~Trott\Curtin$^,$\CAASTRO\myemail,
B.~Pindor\UMelbourne$^,$\CAASTRO,
P.~Procopio\UMelbourne$^,$\CAASTRO,
R.~B.~Wayth\Curtin$^,$\CAASTRO, 
D.~A.~Mitchell\CASS$^,$\CAASTRO, 
B.~McKinley\UMelbourne$^,$\CAASTRO,
S.~J.~Tingay\Curtin$^,$\CAASTRO, 
N.~Barry\UW,
A.~P.~Beardsley\UW$^,$\ASU,
G.~Bernardi\SKASA$^,$\RU$^,$\CfA,
Judd~D.~Bowman\ASU,
F.~Briggs\ANU$^,$\CAASTRO,
R.~J.~Cappallo\Haystack, 
P.~Carroll\UW,
A.~de~Oliveira-Costa\MIT,
Joshua~S.~Dillon\MIT,
A.~Ewall-Wice\MIT,
L.~Feng\MIT,
%B.~M.~Gaensler\USydney$^,$\CAASTRO$^,$\ASTRON, 
L.~J.~Greenhill\CfA,
B.~J.~Hazelton\UW$^,$\UWeSci,
J.~N.~Hewitt\MIT,
N.~Hurley-Walker\Curtin,
M.~Johnston-Hollitt\Victoria,
Daniel~C.~Jacobs\ASU,
D.~L.~Kaplan\UWisc, 
HS Kim\UMelbourne$^,$\CAASTRO,
E.~Lenc\USydney$^,$\CAASTRO,
J.~Line\UMelbourne$^,$\CAASTRO,
A.~Loeb\CfA,
C.~J.~Lonsdale\Haystack, 
%B.~McKinley\UMelbourne$^,$\CAASTRO,
%D.~A.~Mitchell\CASS$^,$\CAASTRO, 
M.~F.~Morales\UW, 
E.~Morgan\MIT, 
A.~R.~Neben\MIT,
Nithyanandan~Thyagarajan\ASU,
D.~Oberoi\Tata, 
A.~R.~Offringa\ASTRON$^,$\CAASTRO, 
S.~M.~Ord\Curtin$^,$\CAASTRO,
S. Paul\RRI,
%B.~Pindor\UMelbourne$^,$\CAASTRO,
J.~C.~Pober\Brown$^,$\UW,
T.~Prabu\RRI, 
%P.~Procopio\UMelbourne$^,$\CAASTRO,
J.~Riding\UMelbourne$^,$\CAASTRO,
N.~Udaya~Shankar\RRI, 
Shiv~K.~Sethi\RRI,
K.~S.~Srivani\RRI, 
R.~Subrahmanyan\RRI$^,$\CAASTRO, 
I.~S.~Sullivan\UW,
M.~Tegmark\MIT,
%S.~J.~Tingay\Curtin$^,$\CAASTRO, 
%C.~M.~Trott\Curtin$^,$\CAASTRO,
%R.~B.~Wayth\Curtin$^,$\CAASTRO, 
R.~L.~Webster\UMelbourne$^,$\CAASTRO, 
A.~Williams\Curtin, 
C.~L.~Williams\MIT,
C.~Wu\UWA,
J.~S.~B.~Wyithe\UMelbourne$^,$\CAASTRO
}

\begin{abstract}
Detection of the cosmological neutral hydrogen signal from the Epoch of Reionization, and estimation of its basic physical parameters, is the principal scientific aim of many current low-frequency radio telescopes. Here we describe the Cosmological HI Power Spectrum Estimator (CHIPS), an algorithm developed and implemented with data from the Murchison Widefield Array (MWA), to compute the two-dimensional and spherically-averaged power spectrum of brightness temperature fluctuations. The principal motivations for CHIPS are the application of realistic instrumental and foreground models to form the optimal estimator, thereby maximising the likelihood of unbiased signal estimation, and allowing a full covariant understanding of the outputs. CHIPS employs an inverse-covariance weighting of the data through the maximum likelihood estimator, thereby allowing use of the full parameter space for signal estimation (``foreground suppression"). We describe the motivation for the algorithm, implementation, application to real and simulated data, and early outputs. Upon application to a set of 3 hours of data, we set a 2$\sigma$ upper limit on the EoR dimensionless power at $k=0.05$~h.Mpc$^{-1}$ of $\Delta_k^2<7.6\times{10^4}$~mK$^2$ in the redshift range $z=[6.2-6.6]$, consistent with previous estimates.
\end{abstract}



%Institutional footnotes (typeset, then rearrange here to be in order)
\Curtintxt
\CAASTROtxt
\myemailtxt
\UMelbournetxt
\CASStxt
\UWtxt
\SKASAtxt
\RUtxt
\CfAtxt
\ASUtxt
\ANUtxt
\Haystacktxt
\MITtxt
\Victoriatxt
\UWisctxt
\USydneytxt
\Tatatxt
\RRItxt
\NRAOtxt
\UWAtxt
\ASTRONtxt
%\Dunlaptxt
\UWeScitxt
\Browntxt

\section{Introduction}
Detection of a neutral hydrogen signal from the Epoch of Reionization (EoR), and estimation of its basic physical parameters, are primary science goals of current and future low-frequency radio telescopes. E.g., Murchison Widefield Array (MWA){\footnote[1]{http://www.mwatelescope.org}} \citep{lonsdale09,tingay13_mwasystem}; Precision Array for Probing the Epoch of Reionization (PAPER){\footnote[2]{http://eor.berkeley.edu}} \citep{parsons10}; the Low Frequency Array (LOFAR){\footnote[3]{http://www.lofar.org}} \citep{vanhaarlem13}; the Long Wavelength Array (LWA){\footnote[4]{http://lwa.unm.edu}} \citep{ellingson09}; and Hydrogen Epoch of Reionization Array (HERA){\footnote[5]{http://reionization.org}}. The neutral hydrogen signal during the EoR corresponds to brightness temperature fluctuations in the gas, and traces conditions within the intergalactic medium (IGM) through a suite of radiative and collisional processes \citep{furlanetto06}. The spatial structure of the signal, as a function of redshift, traces the evolution of the first ionizing sources of radiation in the Universe \citep[e.g.,][]{fialkov14,watkinson14a,pacucci14,pober14,mesinger14,mesinger11}. These studies also predict the signal amplitude to be weak (10s~mK), compared with other sources in the sky (100s~K), and with radiometric noise associated with the internal electronics of the antennas and receiver systems. The detection experiment itself is therefore difficult, and the estimation experiment more so.

Given the current lack of a detection of the signal, and the lack of an instrument with sufficient sensitivity to directly image the brightness temperature fluctuations, research is focused on its statistical detection \citep[][and references therein]{ali15,jacobs14,patil14,dillon13,liu11}. This has the twofold advantage of increasing signal strength by integrating over large areas of the sky, and providing a global statistical indication of the signal (as compared with a local sample obtained by imaging a small patch of sky). The variance (power) of the temperature fluctuations is used as a statistical metric in most studies, with many computing the variance as a function of spatial scale on the sky (the power spectrum). The probability distribution function of the signal is expected to follow a zero-mean Gaussian distribution at early times, and a skewed Gaussian distribution at late times \citep{wyithe07,furlanetto06}. The variance, therefore, captures a lot of the information about the signal. Although the power spectrum has a well-defined mathematical form, factors such as the weakness and complexity of the signal, complexity of the instrumental response function (e.g., frequency- and direction-dependent antenna beam response, bandpass response), and presence of structured, contaminating foregrounds, demand a careful approach to robustly demonstrate that the power spectrum of \textit{cosmological} temperature fluctuations has been measured. Thus, there are a suite of approaches being undertaken by different research groups and instruments \citep{hazelton15,dillon15,chouduri14,patil14,chapman14,jacobs14}. In this work, we describe one of the algorithms being developed for detection of the EoR signal, and computation of a two-dimensional (2D) and spherically-averaged (1D) power spectrum for the MWA. Other papers will describe the other algorithms under development \citep{hazelton15,dillon15,paul15}. Each has a different approach to computing the same metric with the same data, allowing robust cross-referencing and benchmarking of results. \citet{jacobs15} provides a high-level description of the MWA EoR project and algorithms. The key motivations for the estimator we describe here are (1) inclusion of a full instrument and sky description in the statistical estimator; (2) working directly with measured visibility data, where the data description is computationally tractable; and (3) use of a maximum-likelihood estimator to form the optimal estimate with full covariant error analysis.

In Section \ref{sec:mwa} we briefly describe the MWA, and pertinent design properties that affect how the EoR signal is computed. In Section \ref{sec:ps} the power spectrum is formally defined, and recent upper limits on EoR detection using the power spectrum are reviewed. We then introduce the motivation for the CHIPS algorithm and the mathematical basis of its computation. Section \ref{sec:fg} describes the CHIPS approach to foregrounds, and derives the expected signal from foregrounds in the power spectrum parameter space (i.e., `the wedge'). The algorithm is then tested with realistic simulations in Section \ref{sec:sims} and Section \ref{sec:observations} describes the calibration process and the observations used in this work. CHIPS is then applied to MWA data in Section \ref{sec:test}.

Throughout we use a $\Lambda$CDM cosmology with H$_0$=70.4 kms$^{-1}$Mpc$^{-1}$, $\Omega_M$=0.27, $\Omega_k$=0, $\Omega_\Lambda$=0.73 \citep{bennett12}. The discrete Fourier Transform convention used is such that:
\begin{equation}
\tilde{f}_k = \frac{1}{N} \sum_{j=1}^{N} f_j\exp{(-2\pi{i}jk/N)},
\end{equation}
for the forward transform. Vector quantities are expressed with an over-arrow ($\vec{r}$), Fourier-space quantities have tildes ($\tilde{f}$), Fourier-space matrices are boldfaced ($\bf{C}$), and mean values are overlined ($\overline{S}$). The notation, $\tilde{S} \sim \mathcal{CN}(\mu,\boldsymbol{C})$, compactly describes a variable that is statistically distributed as a complex-normal with mean $\mu$ and covariance $\boldsymbol{C}$.

\section{The Murchison Widefield Array}\label{sec:mwa}

The MWA \citep{lonsdale09,tingay13_mwasystem} is an aperture-array low-frequency radio telescope operating in Western Australia in the 80--300~MHz frequency range (instantaneous available bandwidth of 30.72~MHz). Its science themes include EoR detection, discovery and monitoring of radio transients and variables, a southern survey covering 3$\pi$ steradians \citep{wayth15}, and solar and heliospheric science \citep{bowman13_mwascience}. The array consists of 128 tiles, each of which comprises 16 dual-polarization crossed dipole antennas in a regular 4-by-4 grid, with an effective area per tile, $A_{\rm eff}\simeq{21}{\rm m}^2$ (150~MHz).

The antennas are connected to analogue beamformers and the signals correlated within an onsite building. Correlated visibilities in full instrumental polarization flow in real-time to a computing centre located in Perth. The simplicity of the array design and lack of mechanical steering yields a low-cost instrument well-suited to the remoteness and climatic conditions of the desert, but determines the instrumental response to signal from the sky. In particular, the regular grid of dipoles operating as a phased array yields a frequency-dependent beam response with grating lobes (beam nulls), and the analogue beamforming enforces a discretized grid of pointings on the sky where the beam is well-behaved. These design features demand that the position- and frequency-dependent beam response is known and that this knowledge must be incorporated into any estimator (because the beam is the `window' through which the sky signal is observed).

The 30.72~MHz of instantaneous bandwidth is natively captured at 10~kHz spectral resolution, but the standard correlation modes yield 40~kHz channels. Each of these fine channels is contained within 1.28~MHz coarse channels (24 across the band), with substantial attenuation between coarse channels. Typically, two fine channels are flagged with zero data in these nulls between the coarse channels, leading to a bandpass shape with a regular spacing of zeroes. Again, this design feature needs to be correctly accounted for in the analysis, and prevents use of some standard techniques.

\section{Methodology}\label{sec:ps}
\subsection{The power spectrum metric}
The power spectral density (power spectrum) measures the spatial covariance of a signal, integrated over a spatial volume, and corresponds to the Fourier Transform of the two-point correlation function (i.e., the autocorrelation function), $\xi(r)$:
\begin{eqnarray}
P(\vec{k}) &=& \int_{V} \xi(\vec{r}) \exp{(-2\pi{i}\vec{k}\cdot\vec{r})} d\vec{r},\\\nonumber
&=& \int_{V} \langle T(\vec{r}_1)T(\vec{r}_1+\vec{r}) \rangle_{\vec{r}_1} \exp{(-2\pi{i}\vec{k}\cdot\vec{r})} d\vec{r},\\\nonumber
&=& \frac{1}{V} \int_{V} T(\vec{r}_1)T(\vec{r}_1+\vec{r}) \exp{(-2\pi{i}\vec{k}\cdot\vec{r})} d\vec{r}d\vec{r}_1,\\\nonumber
&=& \frac{1}{V} \tilde{T}(\vec{k})\tilde{T}^\ast(\vec{k}),
\end{eqnarray}
where $T(\vec{r})$ is the temperature fluctuation (relative to the mean) at vector position $\vec{r}$, and $\langle\rangle$ denotes an average over positions in the volume, $V$, as a proxy for an ensemble sampling of the distribution. The power spectrum has the units of temperature-squared times volume (${\rm K}^2{\rm Mpc}^3$), and describes the integrated power on a given spatial scale, $\vec{k}$, averaged over the volume, $V$.% For a spherically-symmetric signal, the three $\vec{k}$ vectors can be 

The power spectrum can be computed directly from an image cube, using a three-dimensional Fourier transform of image intensities to find $\tilde{S}(\vec{k})$, and then squaring and normalizing by the cube volume, where $S(\vec{x})$ is the measured brightness temperature in units of Jy$\,$beam$^{-1}$. The conversion from brightness temperature to flux density is given by:
\begin{equation}
S_{\rm Jy} = 10^{26} \frac{2{\rm kT}_K}{\lambda^2}\,\Omega \,\, {\rm Jy},
\end{equation}
is the product of the specific intensity and the observation solid angle, $\Omega$, and k is the Boltzmann constant.

Alternatively, the power spectrum can be computed directly from observed interferometric visibilities. The visibility is the mutual coherence of the two electrical signals detected by the antennas forming a baseline \citep{tms}. In the flat-sky approximation, where the field-of-view (FOV) of the instrument is small such that sky curvature can be ignored, the visibility is identically the Fourier Transform of the product of the sky signal and the beam response. For wide FOV instruments, such as the MWA, this approximation breaks down, and there is a curvature convolving kernel in the measured visibility, in addition to the sky signal and beam response. \citet{thyagarajan15a,thyagarajan15b} studied this effect with MWA data, but current estimators, in general, ignore this effect. The visibility is the natural measurement space of the instrument (the radiometric noise is uncorrelated between visibilities), and will be used in CHIPS to compute the power spectrum directly (without tranforming to and from image space). Curvature sky terms are handled explicitly in CHIPS, and this design choice offers a natural departure from other, image-based power spectrum estimators \citep{hazelton15,paul15,dillon15,patil14,liu11}, but has been used for the angular power spectrum \citep{chouduri14}. A related estimator, the `delay spectrum' \citep[][and references therein]{parsons12,parsons14}, directly Fourier transforms along each visibility's frequency channels, matching a temporal delay with the position of given sky emission relative to the phase centre. This visibility-based estimator approximates the power spectrum on large angular scales, but breaks down for the longer baselines.

The spherically-averaged (1D) power spectrum computes the power on three-dimensional spatial scales, $k=|\vec{k}|$, under the assumption that the statistics of the signal are isotropic and translationally-invariant (the latter being a fundamental assumption of the power spectrum, according to the Wiener-Khinchin theorem). The 1D power spectrum has the advantage of integrating over the largest parameter space, potentially increasing signal detectability. The dimensionless 1D power spectrum, which integrates the total power on a given spatial scale over the volume, is given by:
\begin{equation}
\Delta^2(k) = \frac{k^3}{2\pi^2}P(k).
\end{equation}
PAPER has published the most competitive 1D limits to date, relative to theoretical expectations, yielding (22.4~mK)$^2$ at $z$=8.4 and $0.15h$Mpc$^{-1}<k<0.5h$Mpc$^{-1}$ \citep{ali15}. \citet{patil14} has demonstrated the power of the LOFAR variance statistic, which computes the overall variance in each spectral channel, with simulated data.

It is advantageous, however, to compute the 2D power spectrum as a first step, to discriminate between continuum contaminating foreground sources and the cosmological spectral-line signal. The structure of foregrounds in 2D space is described in Section \ref{sec:fg}, however we comment here that it is substantially different from the cosmological signal. The 2D power spectrum has arguments ($k_\bot,k_\parallel$), where $k_\bot=\sqrt{k_x^2+k_y^2}$, resides in the plane of the sky (angular scales), and $k_\parallel\propto{\eta}$ is the line-of-sight component. Here, $\eta$ is the Fourier dual of frequency, where we use the fact that the observed frequency of the neutral hydrogen spectral line is linearly proportional to distance. Following \citet{morales04} and \citet{mcquinn06}, the transformation between Fourier dimensions and cosmological co-ordinates are:
\begin{eqnarray}
k_\bot &=& \frac{2\pi{|\boldsymbol{u}|}}{D_M(z)},\\
k_\parallel &=& \frac{2\pi{H_0}f_{21}E(z)}{c(1+z)^2}\eta,
\end{eqnarray}
where $D_M(z), H_0, f_{21}, z$ are the transverse comoving distance, Hubble constant, rest frequency of the hyperfine transition ($\sim$1420~MHz), and observation redshift, respectively. $E(z)$ is \citep{hogg99}:
\begin{equation}
E(z) \overset{\underset{\mathrm{def}}{}}{=} \sqrt{\Omega_M(1+z)^2 + \Omega_k(1+z)^2 + \Omega_\Lambda}.
\end{equation}

\subsection{Motivation for CHIPS algorithm}
The high-level design philosophy of CHIPS is to use the data in such a way as to allow for computationally-efficient and robust parameter estimation and error covariance estimation, while retaining maximum information. As such, the following design features have been employed:
\begin{itemize}
\item Calibrated visibilities will be the primary data input --- visibility-space is the natural measurement space of interferometric data, where radiometric (stochastic) noise is uncorrelated and Gaussian distributed with color dependent on the frequency-dependent system temperature. In addition, correlations  due to PSF sidelobes inherent in image-space are naturally accounted for in visibility-space, where the measured information is clearly defined;
\item A Least Squares Spectral Analysis (LSSA) method will be used to compute the line-of-sight transform from frequency to spectral space, not the Fourier Transform \citep{kay93} --- the MWA bandpass has regular missing channels, due to the coarse bandpass edges, and intermittent missing channels due to flagged radio frequency interference \citep[RFI,][]{offringa15}. The Fourier Transform cannot be used for irregular and missing data. LSSA can compute the optimal spectral representation of the data using all of the available information;
\item A maximum-likelihood (ML) estimate of the cosmological signal \citep[a quadratic estimator, ][]{kay93,liu11,dillon13} will be computed (an inverse covariance weighting) --- the ML solution asymptotes to the optimal solution for large quantities of data, and is appropriate for the dataset acquired here. It retains the full information contained within the data, and allows for computation of the full uncertainty covariance matrix. (Note that this is not the only estimator that can be used efficiently to approach this problem.);
\item Foregrounds will be modelled \textit{a priori} and included in the estimator (`foreground suppression') --- as discussed in Section \ref{sec:fg}, foregrounds can be approached in different ways, depending on the degree of knowledge one assumes about them \citep{bonaldi15,chapman14,dillon13,liu11}. Here we take a two-tier approach, where known foregrounds (sky model of point and extended sources) are subtracted from the data coherently, and the remaining signal is treated statistically. CHIPS therefore uses the full Fourier space for estimation (contrast with `foreground avoidance' techniques, where the contaminated regions are excised).
\end{itemize}

\section{Mathematical formalism}
The derivation shown here describes the formation of the power estimate in two steps:
\begin{enumerate}
\item Firstly, it introduces the computation of the coherently-averaged data from measured visibilities, including the incorporation of the frequency-dependent beamshape, curvature terms ($w$-terms), and the transforming of the visibilities in frequency to line-of-sight wavenumber ($\eta$). This step is effectively the visibility gridding (Section \ref{gridding}), and the Fourier-like Transform to ($u,v,\eta$) wavenumber space (achieved using a Least Squares Spectral Analysis - LSSA, in Section \ref{lssa}). The dataset, described by $S(u,v,\nu)$, is a function of location on the $uv$-plane and frequency. Practically, it is obtained by gridding measured visibilities onto the $uv$-plane with the beam kernel (and the $w$-kernel) with a weighting that is a function of the system temperature for that observation and the visibility weight, ($\propto T_{\rm sys}/W$, where $W$ is the visibility weight and corresponds to the number of fine frequency channels averaged to form the visibility). The beam-weighted weights are also accumulated and included in the data covariance matrix, $\boldsymbol{C}$;
\item Secondly, the coherent information forms the power according to a maximum-likelihood estimate, using knowledge about the data covariance matrix (stochastic noise and foreground contaminants, Section \ref{likelihood}).
\end{enumerate}
This process is now described in detail.

%The inverse beam-weighting and LSSA methods are used (Equation \ref{beam_inverse_chips} and Section \ref{lssa}) to form the final, `coherently-averaged data', $S(u,v,\eta)$, which can then be formed into a power. This process is now described.

\subsection{Coherent visibility data: angular modes}
\label{gridding}
We aim to take the large number of measured, calibrated visibilities and grid them onto a regular grid representing the $uv$-plane, in each frequency channel.
The true sky signal is shaped by the antenna primary beam, and the measured visibilities sample a range of modes. Each datum represents information from a region of the $uv$-plane, according to the beam, and we grid a measurement across this region. For an instrument with linear crossed-polarization feeds, $xx$ and $yy$, we model the correlator output as:
\begin{eqnarray}
%\begin{center}
\left[ \begin{array}{c} 
\tilde{V}_{xx} \\ \tilde{V}_{xy} \\ \tilde{V}_{yx} \\ \tilde{V}_{yy} \end{array} \right] = \left[ \begin{array}{c} \tilde{B}_{xx} \\ \tilde{B}_{xy} \\ \tilde{B}_{yx} \\ \tilde{B}_{yy} \end{array} \right] \ast \begin{pmatrix} g_{xx} & g_{xx}\cos{2\psi} & g_{xx}\sin{2\psi} & 0 \\ g_{xy}\Delta_{xy} & -g_{xy}\sin{2\psi} & g_{xy}\cos{2\psi} & i \\ g_{yx}\Delta_{yx} & -g_{yx}\sin{2\psi} & g_{yx}\cos{2\psi} & -i \\ g_{yy} & -g_{yy}\cos{2\psi} & -g_{yy}\sin{2\psi} & 0 \end{pmatrix} \left[ \begin{array}{c} \tilde{I} \\ \tilde{Q} \\ \tilde{U} \\ \tilde{V} \end{array} \right],
%\end{center}
\label{stokes_matrix}
\end{eqnarray}
which maps the Stokes visibilities in sky co-ordinates, $(\tilde{I},\tilde{Q},\tilde{U},\tilde{V})$, to the measurement set in instrumental co-ordinates \citep{hamaker96}. Here we are explicitly describing the relationship in the flat-sky approximation, although the curvature terms are introduced below. The $4\times{4}$ matrix encodes projection effects from the parallactic angle, $\psi$, polarization leakage, $(\Delta_{xy},\Delta_{yx})$, and the direction-independent, antennas-based complex gains, $(g_{xx},g_{xy},g_{yx},g_{yy})$. The convolution accounts for the primary beam shape, $(\tilde{B}_{xx},\tilde{B}_{xy},\tilde{B}_{yx},\tilde{B}_{yy})$. To extract the total intensity, $\tilde{I}$, we could use all four polarization outputs from the correlator. Instead, to reduce computational load, we use only ${V}_{xx}$ and ${V}_{yy}$. This choice is made at the expense of some signal when the zenith angle is large. Failure to treat individual linear feeds will lead to polarization leakage if the visibilities are combined without consideration for beam shape differences \citep{moore13}.

Absorbing the complex gains into the beams, $\vec{B}$, we write the $xx$ and $yy$ visibilities as:
\small
\begin{eqnarray}
\tilde{V}_{xx}(u,v,w) &=& (\tilde{I} + \tilde{Q}\cos{2\psi} + \tilde{U}\sin{2\psi}) \ast \tilde{B}_{xx} \ast \displaystyle\int \frac{\exp{[-2\pi{i}(\boldsymbol{D}\cdot{\boldsymbol{s}})]}}{\sqrt{1-l^2-m^2}}  dldm\\
\tilde{V}_{yy}(u,v,w) &=& (\tilde{I} - \tilde{Q}\cos{2\psi} - \tilde{U}\sin{2\psi}) \ast \tilde{B}_{yy} \ast \displaystyle\int \frac{\exp{[-2\pi{i}(\boldsymbol{D}\cdot{\boldsymbol{s}})]}}{\sqrt{1-l^2-m^2}} dldm,
\end{eqnarray}
\normalsize
where,
\begin{equation}
\boldsymbol{D}\cdot{\boldsymbol{s}} = ul + vm + w(\sqrt{1-l^2-m^2}-1)),
\end{equation}
is the projection of the baseline vector onto the sky co-ordinates, and we have now explicitly included the $w$-terms in the equations through the phase integral (final term).

Using the convolution theorem, the mapping from the underlying sky Fourier representation and the measured visibilities can be written,
\begin{eqnarray}
\tilde{V}_{xx} &=& {\tilde G}_w \ast {\tilde B}_{xx} \ast (\tilde{I} + \tilde{Q}\cos{2\psi} + \tilde{U}\sin{2\psi})\\
\tilde{V}_{yy} &=& {\tilde G}_w \ast {\tilde B}_{yy} \ast (\tilde{I} - \tilde{Q}\cos{2\psi} - \tilde{U}\sin{2\psi}),
\end{eqnarray}
where,
\begin{equation}
{\tilde G}_w \overset{\underset{\mathrm{def}}{}}{=} \displaystyle\int \frac{\exp{[-2\pi{i}(w(\sqrt{1-l^2-m^2}-1))]}}{\sqrt{1-l^2-m^2}} \exp{[-2\pi{i}(ul + vm)]}  dldm,
\end{equation}
explicitly highlights the additional convolution due to curvature terms.
These functions encode the deviation from a strict 2D Fourier transform between sky and visibilities, and the spectral leakage due to the primary beam.

We now expand from considering a single visibility, to a set, and describe these as a vector. In doing so, we extend the convolutions to describe the transform from the underlying sky, $\tilde{I}$, to the full measurement set of data. We can write the discrete convolutions as matrix operations, to encode this transformation, and find the measurements for a set of baselines:
\begin{eqnarray}
\vec{\tilde{V}}_{xx} &=& {\bf G}_w {\bf B}_{xx} (\vec{\tilde{I}} + \vec{\tilde{Q}}\cos{2\psi} + \vec{\tilde{U}}\sin{2\psi})\label{eqn:vxx}\\
\vec{\tilde{V}}_{yy} &=& {\bf G}_w {\bf B}_{yy} (\vec{\tilde{I}} - \vec{\tilde{Q}}\cos{2\psi} - \vec{\tilde{U}}\sin{2\psi}).\label{eqn:vyy}
\end{eqnarray}

We are interested in determining the underlying Stokes I sky distribution at a given angular scale, ${u}_\alpha$, given the measured visibilities, $\vec{\tilde{V}}({u})$. A vector of ($N\times{1}$) measured visibilities, $\vec{\tilde{V}}$, is generated from an underlying ($M\times{1}$) sky distribution, $\vec{\tilde{I}}$, via a ($N\times{M}$) matrix transform, ${G_wB}$. We can re-write and combine Equations \ref{eqn:vxx} and \ref{eqn:vyy} such that:
\begin{equation}
\vec{\tilde{I}}(u,v,\nu) = \frac{1}{2} \left( {\bf B}_{xx}^\dagger {\bf G}_w^\dagger {\bf G}_w {\bf B}_{xx} \right)^{-1} {\bf B}_{xx}^\dagger {\bf G}_w^\dagger \vec{\tilde{V}}_{xx} + \frac{1}{2} \left( {\bf B}_{yy}^\dagger {\bf G}_w^\dagger {\bf G}_w {\bf B}_{yy} \right)^{-1} {\bf B}_{yy}^\dagger {\bf G}_w^\dagger \vec{\tilde{V}}_{yy},
\label{beam_inverse}
\end{equation}
where the square matrix inversion involves inverting an $(M\times{M})$ Hermitian complex-valued matrix, which will be (almost) diagonal for independent modes, $\vec{u}$. Equation \ref{beam_inverse} effectively unwraps the effects of $w$-terms and the polarized beams, before combining $xx$ and $yy$ information to remove the polarized component of the sky signal ($\tilde{Q},\tilde{U}$). The ${\bf G}^\dagger$ matrix serves to distribute the footprint of the visibilities across the $uv$-plane, as dictated by the amplitude of the $w$-term; for large $w$, the Fourier-space beam is effectively broader and the $G$ matrix captures this. In practise, as discussed by \citet{dillon13}, the $uv$-plane needs to be fully-sampled (the matrix full rank) for this matrix inverse to exist (i.e., there needs to be independent information in all modes). For the MWA and using rotation synthesis, this requirement is met for parts of the $uv$-plane, but not in general.

There are multiple approaches to handling this, including computation of the pseudoinverse and simplification of the matrix by considering only diagonal components. The former inverts the matrix in regions where there is sufficient information. The latter effectively ignores the correlations introduced by the finite extent of the beam across the $uv$-plane, and reduces the matrix inverse step to a simple weighting of the data according to the measurements (averaging). It does, however, still encode the distribution of power across multiple modes in the $uv$-plane. The approximation is reasonable given that the matrix is already highly diagonal. In the testing phases of CHIPS, as described here, we will implement the latter:
\begin{equation}
\vec{\tilde{S}}(\vec{u},\vec{v},\nu) = \frac{1}{2} \left( {\bf B}_{xx}^\dagger {\bf G}_w^\dagger {\bf G}_w {\bf B}_{xx} \right)_{\rm diag}^{-1} {\bf B}_{xx}^\dagger {\bf G}_w^\dagger \vec{\tilde{V}}_{xx} + \frac{1}{2} \left( {\bf B}_{yy}^\dagger {\bf G}_w^\dagger {\bf G}_w {\bf B}_{yy} \right)_{\rm diag}^{-1} {\bf B}_{yy}^\dagger {\bf G}_w^\dagger \vec{\tilde{V}}_{yy}.
\label{beam_inverse_chips}
\end{equation}
This introduces an error, for each polarization, with the XX polarization expression:
\begin{equation}
\vec{\tilde{S}}_{XX}(\vec{u},\nu) = \boldsymbol{R}_{XX} \vec{\tilde{I}}_{XX}(\vec{u},\nu).
\end{equation}
Here, the error matrix, $\boldsymbol{R}_{XX}$, is:
\begin{equation}
\boldsymbol{R}_{XX} \overset{\underset{\mathrm{def}}{}}{=} ({\rm diag}(\boldsymbol{B}^\dagger\boldsymbol{B}))^{-1}_{XX}(\boldsymbol{B}^\dagger\boldsymbol{B})_{XX}.
\end{equation}
The same applies to the YY polarization, where the error matrix, in general, is different, depending on the shape of the beams for each telescope pointing. In general, this can invalidate the cancelling of the $Q$ and $U$ linear polarization signals in Equation \ref{beam_inverse}, because there is now the potential for mismatch between the two polarizations. In practise, the XX and YY polarizations are not combined in CHIPS at this point (the results described in this work remain in the two instrumental poarizations), because the current MWA beam model is known to be imperfect. In future, this mismatch will need to be further studied for leakage effects.
%Here we have also separated the full data covariance matrix, $\boldsymbol{C}$, into its thermal noise and foreground covariance, $\boldsymbol{C}_{\rm FG}$, components. The thermal noise, $\sigma(\nu)$ is a function of frequency due to the changing sky temperature with frequency, and practically is computed for each observation by accounting for the measured system temperature for each polarization and the weighting of the visibility.
%Absorbing the error, the covariance matrix describing this signal is then given by:
%\begin{equation}
%\boldsymbol{C}(\nu) = ({\rm diag}(\boldsymbol{B}^\dagger\boldsymbol{B}))^{-1}(\nu)\sigma(\nu)^2 + {C}_{\rm FG}(\nu).
%\end{equation}

%This formalism can support primary beam patterns that differ between antennas and frequencies, and that evolve in time. This allows for changes of the primary beam shape over the course of a pointed observation, and variation in the beam between tiles, as well as frequency-dependent beam shape. Equation (\ref{beam_inverse}) describes a mixing of the measured visibilities across multiple modes. This correlation will become important in the final estimator for the power, describing the degree of covariance between adjacent coherent $uv$ modes due to primary beam effects.

\subsection{Least Squares Spectral Analysis}
\label{lssa}
Upon gridding the data onto the $uv$-plane for each frequency channel, the next step is to estimate the line-of-sight wavenumber information, by transforming the data at each gridpoint from frequency, $\nu$, to $\eta$.
For a complete and regularly-sampled set of $N$ complex-valued discrete datapoints embedded within white Gaussian noise, $\vec{\tilde{S}}(\nu)$, the information contained within a spectral mode, $\eta$, can be obtained using the Discrete Fourier Transform (DFT):
\begin{equation}
\mathcal{S}(u,v,\eta) = \frac{1}{\sqrt{2\pi}}\displaystyle\sum_{j=0}^{N-1} \tilde{S}(u,v,\nu_j) \exp{[-2\pi{i}j\eta/N]},
\end{equation}
where $\eta \in [0,N-1]$. This can be written in matrix formalism, where the Fourier Transform has convenient properties as a Vandermonde matrix;
\begin{eqnarray}
{\vec{\tilde{S}}}(\eta) = \mathcal{F}^\dagger \vec{\tilde{S}}(\nu),\\
\mathcal{F}^\dagger \mathcal{F} = \mathcal{I}.
\end{eqnarray}
Here $\vec{\tilde{S}}(\eta)$ is the complex-valued estimate of the spectral information in mode $\eta$, and $\mathcal{F}$ is a square, Hermitian matrix containing the trigonometric kernel. We have also explicitly dropped the parametrization by $u,v$, because the transform is performed \textit{ for each} point on the $uv$-plane, with the understanding that these parameters are carried along for the computations presented in this section.

When the data have generalized covariance (correlated samples, unequal weightings), the signal in some spectral mode can be estimated using a generalized maximum likelihood Fourier Transform. If the data are distributed such that,
\begin{equation}
\vec{\tilde{S}}(\nu) \sim \mathcal{CN}(\overline{S}(\nu),\boldsymbol{C}),
\end{equation}
the optimal estimate is given by \citep[i.e., the inverse covariance weighting estimator,][]{liu11,dillon15};
\begin{equation}
\vec{\tilde{S}}(\eta) \sim \mathcal{CN}(\mathcal{F}^\dagger \boldsymbol{C}^{-1} \vec{\tilde{S}}(\nu), \mathcal{F}^\dagger \boldsymbol{C} \mathcal{F}),
\end{equation}
where we have used the fact that the Fourier Transform is unitary ($\mathcal{F}^\dagger = \mathcal{F}^{-1}$).
Effectively, this prewhitens the signal (by suppressing data with a lot of noise and removing correlation), and computes the Fourier Transform. The result has noise properties consistent with a prewhitening operation, and with the reordering of the data according to the summing of phased datapoints.

Finally, when the data have incomplete sampling, the above formalism naturally generalizes and is described by Least Squares Spectral Estimation, LSSA. When there are only $M$ sampled points at locations $\nu_j$, within $N$ frequency channels, the generalized least squares estimate is:
\begin{eqnarray}
\vec{\tilde{S}}(\eta) &=& \mathcal{H} \vec{\tilde{S}}({\nu})\\
&\overset{\underset{\mathrm{def}}{}}{=}& \frac{1}{\sqrt{2\pi}}\displaystyle\sum_{j=0}^{M-1} \tilde{S}(\nu_j) \exp{[-2\pi{i}\nu_j\eta/M]},
\end{eqnarray}
where $\eta=[0,...,N_{\rm freq}]$ are evenly-spaced and $N_{\rm freq}<N$. The optimal estimator is:
\begin{equation}
\vec{\tilde{S}}(\eta) \sim \mathcal{CN}((\mathcal{H}^\dagger \boldsymbol{C}^{-1} \mathcal{H})^{-1}\mathcal{H}^\dagger \boldsymbol{C}^{-1} \vec{\tilde{S}}(\nu), (\mathcal{H}^\dagger \boldsymbol{C}^{-1} \mathcal{H})^{-1}),
\label{eqn:lssa}
\end{equation}
where the LSSA operator has been labelled $\mathcal{H}$. Note that now $M<N$ and the $\mathcal{H}$ matrix is not square. In general, this method leads to correlated information between different modes, but this can be reduced by estimating $N_{\rm freq} \lesssim M/2$ modes \citep{vio13}.

\subsection{Likelihood function and ML estimate}
\label{likelihood}
Now we have described the computation of the coherently-averaged (gridded and LSSA) data from the underlying measured visibilities, we can write the likelihood function of the data to compute the power estimates.

We describe the data with a generalized multi-variate Gaussian, with zero mean, and general covariance matrix, ${\bf C}$. This form describes data that are drawn from a statistical distribution of possible universes, with temperature fluctuations that are Gaussian-distributed. This covariance matrix contains all of the terms contributing to the signal in a visibility. The joint likelihood for the vector of complex-valued coherently-averaged data, $\vec{S}(u,v,\eta)$, is given by:
\begin{equation}
L(\vec{\tilde{S}};{\bf C}) = \frac{1}{\pi^N {\rm det}({\bf C})} \exp{[-\vec{\tilde{S}}^\dagger {\bf C}^{-1} \vec{\tilde{S}}]},
\end{equation}
where the complex-valued covariance matrix is;
\begin{eqnarray}
{\bf C} &\overset{\underset{\mathrm{def}}{}}{=}& \left\langle \vec{\tilde{S}} \vec{\tilde{S}}^\dagger \right\rangle\\
&=& {\bf C}_{\rm FG} + {\bf C}_{\rm N} + {\bf C}_{\rm P},
\label{covariance_eqn}
\end{eqnarray}
where,
\begin{equation}
{\bf C}_{\rm P} = \begin{pmatrix}
  p_{11} & p_{12} & \cdots & p_{1N}\\
  p_{12} & p_{22} & \cdots & p_{2N}\\
  \vdots & \vdots & \ddots & \vdots\\
  p_{1N} & p_{2N} & \cdots & p_{NN}
\end{pmatrix},
\label{eqn:cosmo}
\end{equation}
describes the parameters of interest (the mode powers, $p_{\alpha\alpha}$), and the correlations between powers imprinted by the instrument and the experiment (``window functions"). Strictly, the cosmological information forms a diagonal matrix, with uncorrelated power between modes, but the imperfect data correlate contiguous modes. In this implementation of CHIPS, we assume a diagonal cosmological signal matrix, but do encode the correlations between modes in the data by capturing and propagating the covariances introduced by the line-of-sight transform (LSSA). These correlations are important when combining modes together to perform the binning to 1D (see Section \ref{power:lssa}). Note that here we are explicitly identifying the power, in which we are interested, with the likelihood function of the coherently-averaged data. The other terms in equation (\ref{covariance_eqn}) describe the statistical contribution to mode ($u,v,\eta$) from foregrounds (${\bf C}_{\rm FG}$) and measurement (radiometric) noise (${\bf C}_{\rm N}$). In $uv$ space, measurement noise is uncorrelated and Gaussian-distributed (colored Gaussian noise - CGN), where the `color' for a coherently-averaged datum describes the weighting of the noise due to the number of visibilities contributing to that cell. The statistical foreground contribution will be treated in Section \ref{sec:fg}.

The maximum likelihood estimate, obtained by finding the value of a given parameter that maximizes the likelihood function (or minimizes the negative log likelihood), is asymptotically efficient (achieves optimal estimation precision for large datasets). We aim to determine an expression for the parameters of interest (mode powers) in terms of the data and data covariance matrix. The log likelihood function is minimized by differentiating with respect to the parameter, setting to zero, and solving for that parameter. The derivative of the log likelihood for a zero-mean generalized Gaussian is given by:
\begin{equation}
\frac{\partial{\rm ln}L}{\partial{p_{\alpha}}} = {\rm tr}\left( {\bf C}^{-1} \frac{\partial{\bf C}}{\partial{p_{\alpha}}} \right) - \vec{\tilde{S}}^{\dagger}(u,v,\eta) {\bf C}^{-1} \frac{\partial{\bf C}}{\partial{p_{\alpha}}}  {\bf C}^{-1} \vec{\tilde{S}}(u,v,\eta),
\label{eqn:loglike}
\end{equation}
where `tr' denotes the trace of the matrix. For clarity, we now drop the dependence of the data on ($u,v,\eta$), with the understanding that we are working with the data in wavenumber space. Setting Equation \ref{eqn:loglike} to zero, and then using in the first term the identity, $\mathbb{I} = {\bf C}^{-1}{\bf C}$, and expanding the covariance matrix into its constituent parts, we find:
\begin{eqnarray}
\hat{p}_{\alpha} {\rm tr}\left( {\bf C}^{-1}{\bf C}^{-1} \frac{\partial{\bf C}}{\partial{p_{\alpha}}} \right) &\overset{\underset{\mathrm{def}}{}}{=}& {\rm tr}\left( {\bf C}_{\rm P} {\bf C}^{-1}{\bf C}^{-1} \frac{\partial{\bf C}}{\partial{p_{\alpha}}} \right)\\
&=& \vec{\tilde{S}}^\dagger {\bf C}^{-1} \frac{\partial{\bf C}}{\partial{p_{\alpha}}}  {\bf C}^{-1} \vec{\tilde{S}} - {\rm tr}\left( ({\bf C}_{\rm FG} + {\bf C}_{\rm N}) {\bf C}^{-1}{\bf C}^{-1} \frac{\partial{\bf C}}{\partial{p_{\alpha}}} \right),
\label{power_est}
\end{eqnarray}
where $\hat{p}_{\alpha} = \hat{p}_{\alpha\alpha}$ denotes our \textit{estimate} of the power in mode $\alpha$, as described in the diagonal elements of the cosmological power matrix, Equation \ref{eqn:cosmo}. Note here that the first term in the RHS of Equation \ref{power_est} is squaring the coherently-averaged data, and therefore producing the desired power-like quantity.

Therefore, the estimate of the power in mode $\alpha$ is given by computation of:
\begin{equation}
\hat{p}_{\alpha} = \frac{1}{{\rm tr}\left( {\bf C}^{-1}{\bf C}^{-1} \frac{\partial{\bf C}}{\partial{p_{\alpha}}} \right)} \left( \vec{\tilde{S}}^\dagger {\bf C}^{-1} \frac{\partial{\bf C}}{\partial{p_{\alpha}}}  {\bf C}^{-1} \vec{\tilde{S}} - {\rm tr}\left( ({\bf C}_{\rm FG} + {\bf C}_{\rm N}) {\bf C}^{-1}{\bf C}^{-1} \frac{\partial{\bf C}}{\partial{p_{\alpha}}} \right) \right),
\label{estimator_equation}
\end{equation}
where we decouple dependence of the estimator on the mode powers by making the approximation,
\begin{equation}
{\bf C} \approx {\bf C}_{\rm FG} + {\bf C}_{\rm N},
\end{equation}
under the assumption that ${\bf C}_{\rm FG} + {\bf C}_{\rm N} \gg {\bf C}_{\rm P}$.

In practice, the data can be split into two and cross-correlated. This removes thermal noise power from the final power estimate, at the expense of a factor of $\sqrt{2}$ in the final signal-to-noise ratio (in temperature; a factor of 2 in power). Although noise power has been removed from the estimate, thermal noise \textit{uncertainty} remains in the final covariance of the power estimates. Typically, alternate correlator output datasets are processed into alternate coherent data vectors, to ensure uniform $uv$ coverage between the two (this is important for applying the same data covariance matrix to both datasets). Using this scheme, the cross-correlation power formed from coherent data vectors, $_1\vec{S}$ and $_2\vec{S}$, is:
\begin{equation}
\hat{p}_{\alpha} = \frac{1}{{\rm tr}\left( {\bf C}^{-1}{\bf C}^{-1} \frac{\partial{\bf C}}{\partial{p_{\alpha}}} \right)} \left( {_1\vec{S}}^\dagger {\bf C}^{-1} \frac{\partial{\bf C}}{\partial{p_{\alpha}}}  {\bf C}^{-1}  {_2\vec{S}} - {\rm tr}\left({\bf C}_{\rm FG} {\bf C}^{-1}{\bf C}^{-1} \frac{\partial{\bf C}}{\partial{p_{\alpha}}} \right) \right).
\label{estimator_equation_cross}
\end{equation}
The expected values and covariances are derived in the Appendix, and demonstrate that the expected value of the ML estimate yields the cosmological power.

The second term in Equation \ref{estimator_equation_cross} is the foreground power bias. As discussed in a number of recent papers \citep{liu11,dillon13,dillon15}, subtraction of this term requires an accurate model for the foregrounds. Inaccurate subtraction leads to bias in the power. Given the relative power of the foreground and cosmological signal, mis-subtraction will occur when the foreground power is incorrect at the fraction of a percent level. To avoid this, we retain the foreground power bias, and allow the noise covariance to indicate which modes are contaminated (i.e., the second term is set to zero).

\subsubsection{Explicitly including LSSA}
\label{power:lssa}
To connect the LSSA transform with the ML estimate, we can take the expression for the optimal power, Equation \ref{estimator_equation_cross}, which is a function of ($u,v,\eta$), and insert the LSSA estimate of $\vec{\tilde{S}}(u,v,\eta)$ from the underlying gridded data, $\vec{\tilde{S}}(u,v,\nu)$.
To do so, we combine equations \ref{eqn:lssa} and \ref{estimator_equation_cross} to compute the optimal (ML) power estimate:
\begin{equation}
\hat{p}_\alpha = \frac{ (\mathcal{H}^\dagger\boldsymbol{C}^{-1}{_1\overline{S}}(\nu))^\dagger \frac{\partial{\bf C}}{\partial{p_{\alpha}}} (\mathcal{H}^\dagger\boldsymbol{C}^{-1}{_2\overline{S}}(\nu)) } { {\rm tr}\left( (\mathcal{H}^\dagger\boldsymbol{C}^{-1}\mathcal{H}) \frac{\partial{\bf C}}{\partial{p_{\alpha}}} (\mathcal{H}^\dagger\boldsymbol{C}^{-1}\mathcal{H}) \right) }.
\end{equation}
The denominator (the normalization) reduces to:
\begin{equation}
\displaystyle{\sum_i} |A_{i\alpha}|^2,
\end{equation}
where $A_{ij}$ are the elements of the inverse covariance matrix,
\begin{equation}
\boldsymbol{A} \overset{\underset{\mathrm{def}}{}}{=} \mathcal{H}^\dagger\boldsymbol{C}^{-1}\mathcal{H}.
\end{equation}
This then yields the sum over the square of the matrix elements for the normalization. Finally, the vector of power estimates is $\chi^2$-distributed:
\begin{equation}
\hat{\vec{p}}({\eta}) \sim \chi^2\left({\vec{p}} , 2\left( (\mathcal{H}^\dagger\boldsymbol{C}^{-1}\mathcal{H}) (\mathcal{H}^\dagger\boldsymbol{C}^{-1}\mathcal{H}) \right)^{-1}\right),
\label{chi2_dist}
\end{equation}
where the (identity) diagonal cosmological signal matrix, $d\boldsymbol{C}/dp$, has been dropped.

To compute the spherically-averaged (1D) power spectrum from the 2D output, we average in elongated cylindrical shells, allowing for differences in the line-of-sight and angular modes for the cosmological signal in redshift space in the conversion from $u,v,\eta$ to $k$ (Mpc$^{-1}$). The ML estimate of the 1D power is given by weighting the individual 2D mode powers contributing to a cell, $k$, with the inverse of their power covariance matrix. This is the inverse of the covariance term from Equation \ref{chi2_dist}, with:
\begin{equation}
\mathcal{A} \overset{\underset{\mathrm{def}}{}}{=} (\mathcal{H}^\dagger\boldsymbol{C}^{-1}\mathcal{H}) (\mathcal{H}^\dagger\boldsymbol{C}^{-1}\mathcal{H}).
\end{equation}
Thus:
\begin{equation}
P_k = \frac{\displaystyle{\sum_{i\in{k}}}\mathcal{A}_{,i}\vec{p}}{{\rm tr}(D^\dagger\mathcal{A}_{,i}D)},
\end{equation}
where the covariance matrix of data is used quadratically to reflect that we are now working to optimally average \textit{power}, and $D$ is a binning matrix which reduces the full 2D space to the subspace spanned by $i\in{k}$. The numerator terms here come from performing the matrix multiplication from Equation \ref{chi2_dist}, and including all of the covariances between 2D modes, $i$, contributing to binned mode $k$. The denominator provides the variances and covariances between the 1D $k$ modes, encoding the degree of uncertainty on each parameter and the relationship between estimates. The syntax, $_{,i}$, denotes the matrix, $\mathcal{A}$, as formed from projecting onto the bin, including parameter covariances.

\section{Building the data covariance matrix}
\subsection{Foregrounds}\label{sec:fg}
Dealing with foreground contamination can be approached in different ways, all of which rely on the basic distinction that foreground sources are continuum emitters, and therefore have a smooth spectrum over a small bandwidth, and the cosmological signal is a spectral line with structure reflecting the brightness temperature at different spatial depths. This distinction fundamentally dissociates angular modes from line-of-sight modes, and is the reason for operating with the 2D power spectrum. In 2D, the power from flat spectrum foregrounds that could be observed through a perfect instrument (no frequency-dependent instrumental response, no spectral incompleteness, and complete sampling of the Fourier plane), would be contained within the DC ($\eta=0$) mode of $k_\parallel$, because they add a simple amplitude across frequency. For a real instrument and non-flat sources, foregrounds occupy a broader region of parameter space, colloquially termed the `wedge', because of its broad wedge-like structure in ($k_\bot,k_\parallel$) space. The wedge has been well-studied \citep{datta10,trott12,morales12,vedantham12,hazelton13,thyagarajan13,parsons12}. It results from the incomplete sampling of the Fourier modes and/or spectral modes by an interferometer, and the migration of the angular mode sampled by a given baseline as a function of observation frequency (`mode-mixing'). A mathematical derivation of the expected structure of the wedge for the MWA is included in this section.

Broadly, there are two primary foreground treatment design options employed in the literature: (1) `avoidance', where it is attempted to contain foregrounds to a region of parameter space, and ignore this region \citep[e.g., PAPER's approach,][]{parsons12,jacobs14}; (2) `suppression', where the foreground contribution to the signal is modelled (using either an \textit{a priori} source model, non-parametric methods, or a data-driven model) and the estimator uses this knowledge to suppress contaminated information. The latter option includes the non-parametric fitting of low-order polynomials \citep{bowman09}, and the more sophisticated Component Analysis techniques discussed by \citet{chapman14,chapman12,bonaldi15}, and references therein. It also includes model- and data-driven (parametric) statistical descriptions of the sky itself, and incorporation of that information \citep{liu11,dillon13,dillon15}. These approaches all have their own advantages and disadvantages. Broadly, non-parametric approaches are simple to implement, but, requiring no input physical knowledge, can destroy cosmological information and retain foreground signal. Typically, the metric for an adequate solution is not well-defined. Conversely, parametric models are designed to include as much physical information about the foregrounds and cosmological signal as is available. However, they can be computationally difficult to implement, are limited by our knowledge of the sky at low radio frequencies, and can also destroy information if the models are incorrect. Surveys such as the MWA's GLEAM \citep{wayth15} and LOFAR's MSSS \citep{heald15}, will be crucial for improving our understanding.

Ultimately, the key to a robust estimator is to have an understanding of the limitations and biases of one's method, and incorporate that knowledge into the methodology and results. CHIPS aims to do this, and handles foregrounds using a two-tiered approach: (1) known foregrounds (sky model of point and extended sources, with updated calibration solutions and ionospheric corrections) are subtracted from the data coherently (visibility data) using the Real-Time System (RTS) calibration (see Section \ref{sec:observations}); (2) remaining signal is handled statistically, using an \textit{a priori} model for the expected distribution and spectral structure of sources (CHIPS). We then use a perturbative calculation to ascertain our degree of confidence in the model \citep{liu15}.

We initially employ a two-component statistical foreground model, consisting of extragalactic point sources and Galactic synchrotron, described here. The component describing the Galactic Plane is omitted. The Galactic Plane is more difficult to describe statistically, with definite sky position and skewed statistics. We leave this as an open problem for future work.

\subsubsection*{Point source covariance matrix}\label{sec:pt_source}
We consider the additional noise-like signal contained within a visibility due to unmodelled point sources present within the primary beam (where the Poisson-distributed number of sources within any differential patch of sky yields the variant, noise-like signal). To compute the contribution of these sources, we perform the following calculation. The number of sources within a small area of the sky is assumed to be Poisson-distributed. For a Poisson distribution, the variance is equal to the mean.% For pedagogy, we consider a range of models, each with increasing complexity.
\begin{enumerate}
\item Calculate the Poisson noise due to the random number of sources within a small patch of sky and a small range of source flux density ($N(S,S+dS;l,l+dl;m,m+dm)$);
\begin{equation}
N(S,S+dS;l,l+dl;m,m+dm) = \frac{dN}{dS}dSdldm,
\end{equation}
where $dN/dS$ is the source density per unit flux density, and is given parametrically by,
\begin{equation}
\frac{dN}{dS}(\nu) = \alpha \left( \frac{S_{\rm Jy}}{S_0}\right)^{-\beta} {\rm Jy^{-1} sr^{-1}}.
\end{equation}
For this work we use the number counts of \citet{intema11}, with $\alpha=4100{\rm Jy}^{-1}{\rm sr}^{-1}$, and $\beta=1.59$ at 150~MHz.

\item Compute the variance on a visibility measurement due to the Poisson noise from a differential patch of sky at $(l^\prime,m^\prime)$;
For $N$ sources with flux density $S(\nu)$ located at sky position $(l^\prime,m^\prime)$ within the beam $B(l,m,\nu)$, the mean visibility is given by;
\begin{equation}
\left\langle V(u,v) \right\rangle = NSB(l^\prime,m^\prime,\nu) \exp{[-2\pi{i}(ul^\prime + vm^\prime)]}.
\end{equation}
%and the variance is;
%\begin{equation}
%\Sigma_{V(u,v)} = NS^2 \left( \frac{\nu}{\nu^\prime} \right)^{-\gamma}  B(l^\prime,m^\prime,\nu)^2,
%\end{equation}
%where, $S(\nu) = \left( \frac{\nu}{\nu^\prime} \right)^{-\gamma}$, contains the spectral index of the sources, and
%$B(l,m,\nu) = B(l,m;\nu)\Gamma(\nu)$,
%is the product of the frequency-dependent spatial beam and any bandpass filter applied \textit{a posteriori} to the spectral channels \citep[e.g., Blackman-Nuttall,][]{thyagarajan13}.
%The covariance between baselines is given by:
%\begin{equation}
%\boldsymbol{C}_{uu^\prime} = NS^2B(l^\prime,m^\prime,\nu)^2 \exp{-2\pi{i}[(u-u^\prime)l^\prime + (v-v^\prime)m^\prime]}.
%\end{equation}

\item Compute the total variance within a visibility from all sources (in the simple model where there is no source clustering, covariance matrices sum because sources are independent);
%The mean number of sources in the sky is given by;
%\begin{equation}
%N_{\rm Tot} = \iiint N(S,S+dS;l,l+dl;m,m+dm)dldm = \iiint \frac{dN}{dS}dSdldm,
%\end{equation}
the total covariance between two visibilities on the same baseline (different frequencies) can be computed by considering the response of the instrument at different frequencies, for a fixed location in the $uv$-plane. The source number density, primary beam, and $uv$-sampling all change, with the latter (the physical source of the wedge) being encapsulated in the term $f_\nu=(\nu^{\prime\prime}-\nu^\prime)/\nu_{\rm low}$, which produces the frequency phase-wrapping that yields the wedge feature in power. Here $\nu_{\rm low}$ is the lowest measurement frequency channel, and $\nu^\prime$ and $\nu^{\prime\prime}$ denote different frequencies within the band. The covariance is\footnote{This expression simplifies for a circularly-symmetric beam, where the 2D Fourier Transform can be written as a 1D Hankel Transform:
\begin{equation}
\boldsymbol{C}_{\rm PS} = \frac{\alpha}{3-\beta} \left( \frac{\sqrt{\nu^{\prime\prime}\nu^\prime}}{\nu_{\rm low}} \right)^{-\gamma} \frac{S_{\rm max}^{3-\beta}}{S_0^{-\beta}} \int_0^\infty B(l;\nu^{\prime\prime})B(l;\nu^\prime) J_0\left({2\pi(ul)(\nu^{\prime\prime}-\nu^\prime)}\right) ldl \hspace{0.3cm} {\rm Jy^2}.
\end{equation}
For a top-hat beam truncated at the horizon ($l=1$), the foreground covariance has a simple form, which is similar to a sinc function:
\begin{equation}
\boldsymbol{C}_{\rm PS} = \frac{\alpha}{3-\beta} \left( \frac{\sqrt{\nu^{\prime\prime}\nu^\prime}}{\nu_{\rm low}} \right)^{-\gamma} \frac{S_{\rm max}^{3-\beta}}{S_0^{-\beta}} \frac{J_1(2\pi{u}f_\nu)}{f_\nu{u}},
\end{equation}
where $f_\nu = (\nu^{\prime\prime}-\nu^\prime)/\nu_{\rm low}$ and $u=|x|\nu_{\rm low}/c$. Similarly, a frequency-dependent Gaussian-shaped beam is often a reasonable approximation to the beam shape;
\begin{equation}
B(l;\nu) \propto \exp{[-l^2/\sigma^2]},
\end{equation}
where,
\begin{equation}
\sigma(\nu) \simeq \epsilon{c}/\nu{D},
\end{equation}
and $\epsilon\simeq{0.42}$, and $D\simeq{4}{\rm m}$ are the scalings from an Airy disk to a Gaussian width, and the tile diameter, respectively.
Using this, and the Hankel Transform of a Gaussian, yields;
\begin{equation}
\boldsymbol{C}_{\rm PS} = \frac{\alpha}{3-\beta} \Gamma(\nu^{\prime\prime})\Gamma(\nu^\prime) \left( \frac{\sqrt{\nu^{\prime\prime}\nu^\prime}}{\nu_{\rm low}} \right)^{-\gamma} \frac{S_{\rm max}^{3-\beta}}{S_0^{-\beta}} \frac{\pi{c^2}\epsilon^2}{D^2}\frac{1}{\nu^{{\prime\prime}2} + \nu^{\prime{2}}} \exp{\left( \frac{-u^2c^2f_\nu^2\epsilon^2}{4(\nu^{{\prime\prime}2} + \nu^{\prime{2}})D^2} \right)},
\end{equation}
where $\Gamma$ indicates the use of a frequency taper function to reduce spectral leakage (e.g., a Hanning Window, as used here).
For a real, frequency-dependent MWA beam, the foreground covariance can be computed numerically.};
\begin{eqnarray}
\boldsymbol{C}_{\rm PS} &=& \iiint S^2 \left( \frac{\sqrt{\nu^{\prime\prime}\nu^\prime}}{\nu_{\rm low}} \right)^{-\gamma} B(l,m;\nu^{\prime\prime}) B(l,m;\nu^\prime) \frac{dN}{dS}dSdldm \\
&=&  \frac{\alpha}{3-\beta} \left( \frac{\sqrt{\nu^{\prime\prime}\nu^\prime}}{\nu_{\rm low}} \right)^{-\gamma} \frac{S_{\rm max}^{3-\beta}}{S_0^{-\beta}} \iint B(\vec{l};\nu^{\prime\prime})B(\vec{l};\nu^\prime) \exp{[-2\pi{i}(\vec{u}\cdot\vec{l})f_\nu]} dldm \hspace{0.3cm} {\rm Jy^2},
\label{foreground_covariance}
\end{eqnarray}
where $S_{\rm max}$ is the brightest unmodelled source in the field (the peeling limit), here taken as 1~Jy.
\end{enumerate}

\subsubsection*{Galactic synchrotron covariance matrix}\label{sec:galsyn}
Galactic synchrotron emission, from electrons spiralling along our Galaxy's magnetic field lines, produces emission on large scales. We use the parameters of \citet{jelic08,jelic10} to motivate our model, which also includes the effects of the MWA primary beam and instrumental chromaticity.

\begin{figure}[t]
\centering
\subfigure[Beam-convolved Galactic synchrotron component.]{
\includegraphics[width=.4\textwidth]{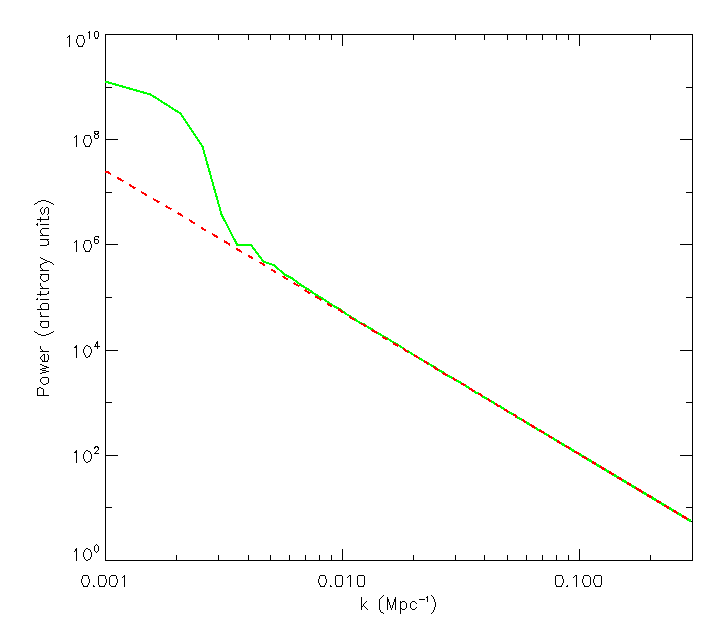}
}
\subfigure[Foreground model.]{
\includegraphics[width=.45\textwidth]{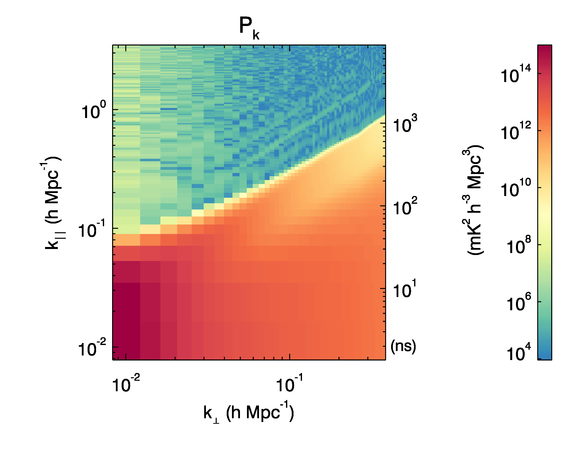}
}
\caption{(a) Intrinsic (red, dashed) and beam-convolved (green, solid) power spectra for the Galactic synchrotron model. (b) Power spectrum for the input foreground covariance model, including contributions from extragalactic point sources and Galactic synchrotron.}
\label{fig:gs_ps}
\end{figure}

The intrinsic temperature power spectrum is modelled as:
\begin{equation}
\langle \Delta{T}^2_{\rm GS}\rangle(u,\nu) = (\eta{T_B})^2\,\left(\frac{u}{u_0}\right)^{-2.7}\left(\frac{\nu}{\nu_0}\right)^{-2.55}\,\,{\rm K^2},
\end{equation}
where $\eta=0.01$ is the fluctuation level relative to the uniform brightness temperature, $T_B=253{\rm K}$, $u_0 = 10$ wavelengths, and $\nu_0=100$~MHz is the reference frequency. The covariance matrix is a function of the parameter, $u$, to reflect that we expect the statistics to be rotationally-invariant. The apparent steep spectral index in temperature units is flattened once converting to flux density (integrated) units for a given instrument, such that the intrinsic power is:
\begin{eqnarray}
P_{\rm GS}(u,\nu) &=& \left(\frac{2k}{\lambda^2}\right)^2\Omega(\eta{T_B})^2\,\left(\frac{u}{u_0}\right)^{-2.7}\left(\frac{\nu}{\nu_0}\right)^{-2.55}\,\,{\rm Jy^2}\\
&=& \left(\frac{2k}{\lambda}\right)^2\frac{1}{A_{\rm eff}}(\eta{T_B})^2\,\left(\frac{u}{u_0}\right)^{-2.7}\left(\frac{\nu}{\nu_0}\right)^{-2.55}\,\,{\rm Jy^2}\\
&=& \left(\frac{(2k)^2}{A_{\rm eff}}\right)(\eta{T_B})^2\,\left(\frac{u}{u_0}\right)^{-2.7}\left(\frac{\nu}{\nu_0}\right)^{-0.55}\,\,{\rm Jy^2}.
\end{eqnarray}
The truncation of the spectrum due to the primary beam can be obtained via convolution (via a Fourier transform), and Figure \ref{fig:gs_ps}(a) shows the intrinsic and convolved power spectra.
Here, the Fourier Transform is used (rather than LSSA) because the plot uses complete and regularly-sampled spacing for display purposes. The actual covariance contains the same incompleteness as the data, when implemented in CHIPS. Note that the primary beam convolution is frequency-dependent, generating complex structure at small wavenumbers.

Finally, the instrumental chromaticity is included in a similar manner as for the point source model, yielding the following model for the spectral covariance matrix:
\begin{equation}
\boldsymbol{C}_{\rm GS}(\nu,\nu^\prime;u) = 2\pi \sqrt{P^\prime_{\rm GS}(u,\nu)P^\prime_{\rm GS}(u,\nu^\prime)} \int J_0\left({2\pi(ul)f(\nu)}\right) ldl \hspace{0.3cm} {\rm Jy^2},
\end{equation}
where $P^\prime$ denotes the beam-convolved power.

\subsubsection{Total covariance matrix}\label{sec:total_cov}
The two foreground components are summed to form the complete foreground contribution to the data covariance matrix, as observed by the instrument. After performing a spectral transform to $\eta$-space, and converting to cosmological units, the final total foreground contribution to the power spectrum is shown in Figure \ref{fig:gs_ps}(b) (coarse bandpass missing channels are omitted for this plot, for clarity). Note the presence of coherent streaks of emission beyond the foreground wedge region, but parallel to it. This is caused by spectral structure in the window taper function (in this example, a Blackman-Nuttall taper is used) interacting with the non-smooth shape of the MWA primary beam. Note also the apparent brightening towards the edge of the wedge region. Naively, the wedge shape in the $k_\parallel$ direction is dictated by the primary beam (for a uniform foreground brightness distribution), suggesting that the edge should be attenuated. However, as predicted in \citet{thyagarajan13} and demonstrated in \citet{thyagarajan15b}, wide-field curvature effectively compresses the sky, increasing the density of emission at the edge of the wedge.

\subsection{Thermal noise}
\label{sec:thermal}
The final component of the data covariance corresponds to the thermal (measurement, radiometric) noise. This is the measurement uncertainty on each visibility due to the finite number of data samples (information) that contribute to it. We are effectively estimating the sky signal with a fixed amount of information, which depends on the signal strength and the number of samples. The former is set by the system temperature, which is dominated by sky temperature at low frequencies, and the latter is set by the bandwidth and sampling time for each visibility. For a single polarization,
\begin{equation}
\sigma = 10^{26} \frac{2\rm{k}\rm{T_{sys}}}{A_{\rm{eff}}}\frac{1}{\sqrt{\Delta\nu\Delta{t}}} \hspace{3mm} {\rm Jy}.
\end{equation}
When gridding the visibilities onto the $uv$-plane, as described in Section \ref{gridding}, the noise decreases coherently (with square-root improvement). This noise reduction therefore follows the evolution of sampled points in the $uv$-plane for a nightly track of the EoR field. Identical observations on subsequent nights yields the same coherent reduction in power. The thermal noise contribution to the power spectrum therefore maps the distribution of points in the $uv$-plane for a nightly track, thereby reflecting the array configuration.

\section{Simulations}\label{sec:sims}
To test the estimator we generated a set of end-to-end noise-free simulations of a single 2-minute snapshot of visibilities, and passed them through the pipeline, using a power-law input power spectrum. The aims of these simulations were to verify that the slope and normalization were unbiased. We chose the amplitude of the spectrum arbitrarily ($A=1\,{\rm K}^2$), and the index was set to $n=-1$, where $P(|k|)=A\,k^{n}\,{\rm K}^2$. To produce realistic visibilities, we included the following in the simulations: (1) The full frequency-dependent primary beamshape of the instrument; (2) The actual $uv$ distribution of a zenith-pointed observation; (3) Curvature of the sky ($w$-terms).

The simulations were produced by starting with an image-cube Gaussian random field of brightness temperature fluctuations ($l,m,\nu$), performing a 3D Fourier Transform to $k$-space, multiplying by the square-root of the input power spectrum, and inverse Fourier Transforming back to the real space-frequency cube. The data were then multiplied by the frequency-dependent beam shape, and regridded to a curved co-ordinate system. Finally, the $uv$-plane for each frequency channel was generated by performing a 2D Fourier Transform to $(u,v,\nu)$-space. This cube formed the underlying data from which the visibilities were sampled according to the baseline distribution of the MWA \citep{beardsley13}.% During the CHIPS processing, a Blackman-Nuttall window was applied to the data in the spectral dimension, to fully reproduce the normal processing of data. No noise was added to the data.

\begin{figure}
\epsscale{0.8}
\plotone{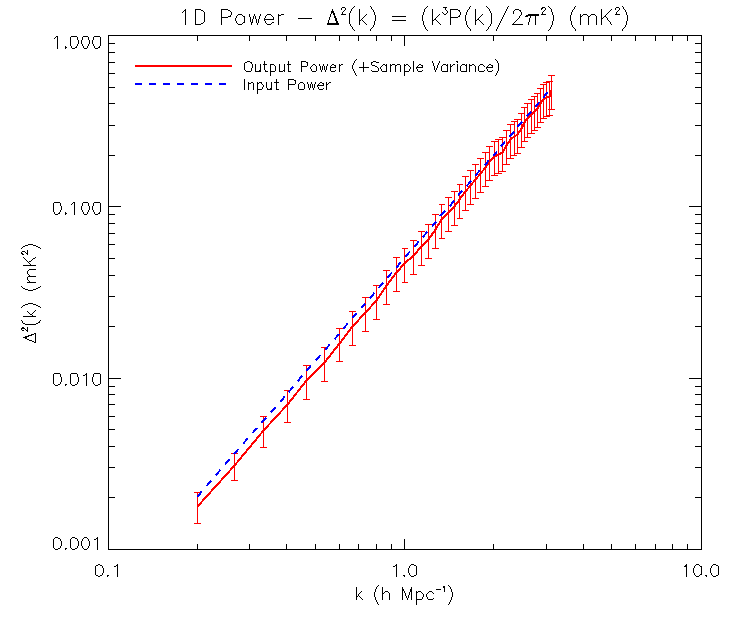}
\caption{Spherically-averaged power spectrum of simulated power spectrum (red solid), and input spectrum (blue dashed). The error bars quantify 2$\sigma$ uncertainties due to sample variance. 
%See the electronic edition of the Journal for a color version 
%of this figure.
\label{fig:sims}}
\end{figure}

Figure \ref{fig:sims} displays the computed spherically-averaged power spectrum and input power spectrum, showing agreement between the two within a few percent.% There is a slight systematic deficit of power observed at small $k$. The discrepancy at the reference wavenumber ($k=1$~Mpc$^{-1}$) is $<$10\%. 

\section{The Real-Time System and Observations}\label{sec:observations}
Data are calibrated using the MWA Real-Time System \citep[RTS,][]{mitchell08,mitchell15}. The RTS has been specifically designed to calibrate MWA data at high cadence, using the full frequency- and direction-dependent beam response, ionospheric modelling and correction, and multi-source in-field calibration using point and extended source models.% \citep{procopio15}. %Calibrated visibilities are available with frequency and temporal resolution of $\Delta\nu=40$~kHz and $\Delta{t}=2$~seconds, respectively. \textbf{[BP:I still think this sentence is superceded by the last sentence of the following paragraph]}

% Reference for karemi http://adsabs.harvard.edu/abs/2013MNRAS.430.1457K
% reference for offringa http://adsabs.harvard.edu/abs/2010ascl.soft10017O

MWA EoR observations used in this work were collected in 112s integrations during which the MWA correlator outputs raw visibilities every 0.5s. Data are used from the MWA EOR0 field, centered at RA=0h, Dec.=$-$27 degrees. This sky position is chosen to avoid the Galactic plane and yield a relatively cold sky.
Radio frequency interference was detected and excised using AOFlagger \citep{offringa12}. Each such observation is independently calibrated by the RTS in a two stage process. First, the entire observation is used to determine the direction-independent Jones matrices (complex gains) of each MWA tile by fitting the observed visibilities to a model which consists of the 1000 apparently\footnote{As attenuated by the primary beam.} brightest sources which lie within 20 degrees of the pointing center. This model is constructed from catalogues of known radio sources using the PUMA algorithm \citep{line15}\footnote{In addition to the positional matching, PUMA uses entries from known catalogues and fits a power law to these data points to assess the reliability of the matching. 
In practise, each source in the catalogue almost always has more than one flux entry. Contiguous entries are then used in the RTS to compute the spectral index for that frequency band. 
The relatively large frequency intervals between the catalogue entries enforce a generally smooth behaviour of the sources in the band considered while preserving the overall SEDs features.}. This method of combining faint sources into a single high S/N calibrator is similar to source clustering as proposed by \citet{kazemi13}, although in this case we only construct a single compound source. Following this averaged calibration stage, the sky model sources are subsequently individually passed through the RTS Calibrator Measurement Loop (CML). For each source within the CML, i) the model visibilities of all other calibrator sources are subtracted, ii) an ionospheric offset and gain\footnote{Ostensibly, this term is intended to account for ionospheric attenuation. However, at present it subsumes and it is likely dominated by errors in the interpolated catalogue source fluxes.} term is measured by fitting a phase ramp to the visibilities when rotated towards the catalogue position of the calibrator, and iii) for the brightest sources, direction-dependent (DD) corrections to the antenna gains are fitted to the residual visibilities following ionospheric correction. In practice, a combination of limited processing time, S/N, and available degrees of freedom limit the number of sources which can receive this full direction-dependent treatment. In this work, five sources are treated as full direction-dependent calibrators and the rest are only updated for ionospheric effects as above. For the EOR0 field, which has relatively few bright sources, the DD calibrators have flux densities $\sim$~10$-$20 Jy. Once the CML model has converged, the best-fit model of each calibrator source is subtracted from the visibilities. The RTS is parallelized over frequency, so that each of the 24 coarse channels (1.28~MHz) is effectively independently calibrated. One exception to this are the ionospheric offset fits which are fitted to a $\lambda^2$ dependence over the entire bandwidth. The second calibration and subtraction step is performed on an 8 second cadence in order to resolve in time ionospheric fluctuations. Note that for sources subtracted during this step, the requirement that they must lie within 20 degrees of the pointing center is removed so that sufficiently bright sources can be subtracted from anywhere in the sky. As a result, the lists of sources used in the two calibration stages are not identical. Subtracting 1000 sources corresponds to a subtraction threshold of $\sim$~350~mJy at the center of the beam. All calibration operations are performed at 40 kHz frequency resolution but the residual visibilities are averaged to 80 kHz for power spectrum estimation. Hence, the output from the RTS to the power spectrum estimation module are calibrated, residual visibilities averaged to 80~kHz and 8s.  

\begin{figure}
\epsscale{0.8}
\plotone{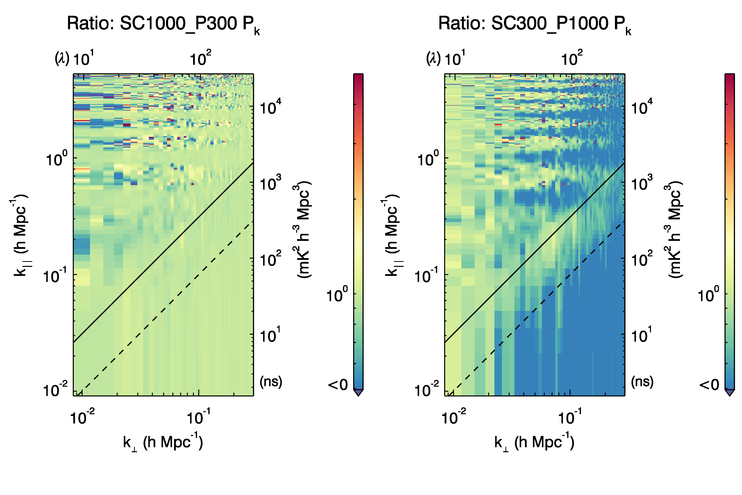}
\caption{Ratio of the power in two calibration and peeling strategies, relative to a standard strategy of self-calibrating with 300 sources and peeling 300 sources, as described in the text. Self-calibrating on more sources (left) produces a more well-calibrated dataset and yields improvement in the non-foreground dominated regions of the power spectrum. Leaving the calibration the same but peeling more sources removes power from the foregrounds, thereby yielding lower power in the wedge and coarse bandpass harmonics. The diagonal dashed and solid lines refer to where flat-spectrum foregrounds would lie if contained within the main lobe of the primary beam, and the horizon, respectively. \label{fig:calibration_ratios}}
\end{figure}

Different calibration and peeling strategies were tested to find the optimal calibration settings for these data. Beyond the usual image-quality and calibration-stability metrics employed to assess calibration performance, we also used the power spectrum to assess the impact of different strategies. Three such modes were (1) Self-calibrate with 300 sources, peel 300 sources; (2) Self-calibrate with 1000, peel 300 sources; and (3) Self-calibrate with 300 sources and peel 1000 sources. Figure \ref{fig:calibration_ratios} shows the ratio of cross power for strategies (2) and (3), relative to strategy (1) (which we expect to leave the most power in the power spectrum), for the fifteen zenith-only observations of the data used in this work.
Calibrating on more sources leads to a more precise calibration solution, and improves the regions of interest for EoR science. Peeling additional sources reduces power in the wedge, as expected for removing signal power. An interesting feature of removing more wedge power is the reduction in power at higher $k_\parallel$, where copies of the wedge caused by harmonics introduced by the regular coarse bandpass missing channels leak power beyond low $k$ modes. Hence, peeling more sources allows a reduction in foreground power throughout the 2D power spectrum parameter space.
Ultimately, a strategy self-calibrating with 1000 sources and peeling 1000 sources was employed for the work presented in this paper.

\section{Results}\label{sec:test}

\subsubsection*{Application to data}
The set of high-band EoR data were taken from a single night of observations during the first semester of MWA EoR observing (2013, August 23). These data were chosen to test, refine and compare different calibration and analysis methodologies, as described in \citet{jacobs15}. The original dataset consists of 160 112-second observations. These data were refined to a final set of 94 observations, removing pointings that were $>25$~degrees from zenith, and a pointing heavily affected by Galactic emission in the sidelobes. As described in Section \ref{sec:observations}, the data were calibrated and provided as 8~second visibilities, yielding 14 timesteps per observation, and seven per interleaved dataset per observation (temporally-interleaved data are used to compute the cross power spectrum). After calibration, we performed a $uv$ cut at a maximum distance of 300 wavelengths at the lowest frequency, within which the EoR signal is expected to fall, and used the full bandwidth dataset for the initial analysis (30.72~MHz).

The cross power spectra are produced with CHIPS both with, and without, the foreground covariance matrix included in the estimator. Without foreground covariance, the data are weighted purely by the system temperature for that observation, the relative weight of the visibility (determined from the number of 10~kHz channels contributing), and the $uv$-sampling of the instrument, and corresponds to the most straight-forward application of a power spectrum analysis. With foreground covariance, the data are down-weighted according to the modelled and measured noise and signal contamination within the data.

\begin{figure}
\epsscale{0.8}
\plotone{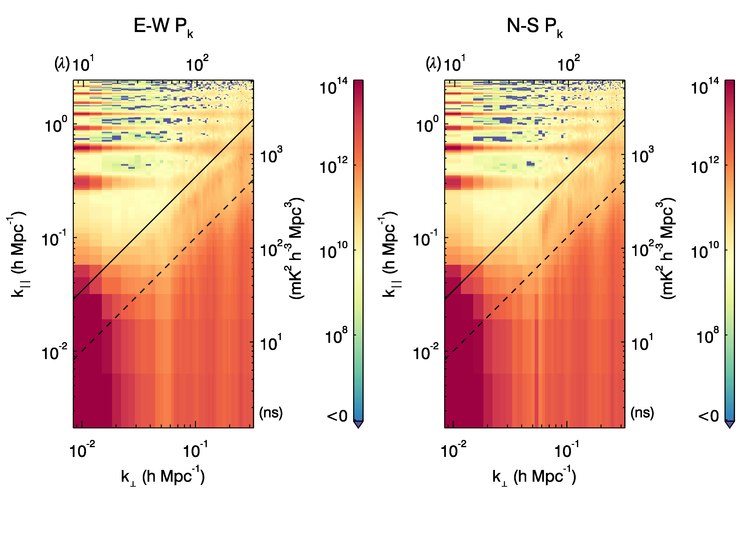}
\caption{E-W (left) and N-S (right) cross power spectra when foregrounds are \textit{excluded} from the estimator. Both polarizations demonstrate foreground power consistent with smooth point sources and large-scale Galactic emission. The N-S polarization has additional power beyond the theoretical edge of the main primary beam lobe (into the region approaching the horizon) and corresponds to Galactic emission rising or setting during the observations. \label{fig:2d_power}}
\end{figure}

\begin{figure}
\epsscale{0.8}
\plotone{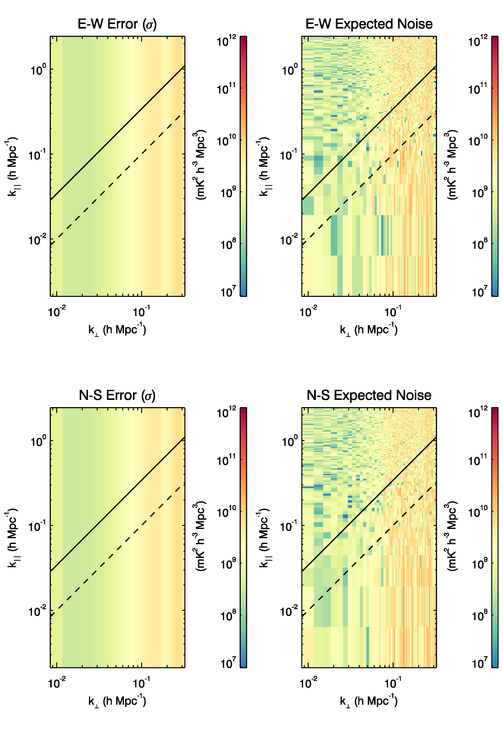}
\caption{Error and expected noise computed from the input visibility weights and a system temperature of 240~K at 170~MHz, for both instrumental polarizations. \label{fig:2d_noise}}
\end{figure}

\begin{figure}
\epsscale{0.8}
\plotone{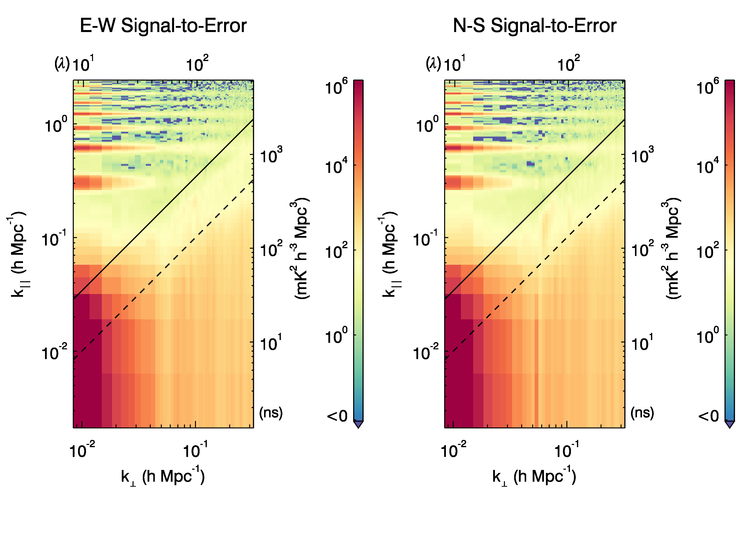}
\caption{Signal-to-error ratio (ratio of cross power to errors) for both polarizations where the errors term only includes the thermal noise contribution to the estimator (no foreground contribution). As expected when the foreground contribution is excluded from the estimator, the foreground-dominated wedge and coarse bandpass harmonics shows high signal-to-error ratio detections of signal. \label{fig:2d_snr}}
\end{figure}

Figure \ref{fig:2d_power} shows the instrumental North-South and East-West cross power polarizations for the full bandwidth.
The solid and dashed diagonal lines correspond roughly to the expected locations of foregrounds within the main lobe of the MWA primary beam and the horizon, respectively. There is clear foreground contribution within the expected region. Note that this wedge emission is consistent with the input foreground model, as shown in Figure \ref{fig:gs_ps}. There is also the clear imprint of the MWA's $uv$-sampling function, which has few very short and few very long ($k>0.08$h.Mpc$^{-1}$) baselines. The copies of the wedge are visible at regular intervals in $k_\parallel$, corresponding to the comb-like bandpass sampling function of the MWA. The errors and expected noise, and the measured signal-to-error ratio can also be computed (Figures \ref{fig:2d_noise} and \ref{fig:2d_snr}).
The error is computed using Equation \ref{chi2_dist} for the thermal-noise only data covariance matrix, as considered for the estimator (foreground data covariance contribution is omitted here). The expected thermal noise is computed by gridding visibilities with a thermal noise contribution given an average system temperature of $T_{\rm sys}=240$~K, and propagating through the same estimator. This system temperature was estimated by matching the observed noise (obtained through the difference in the even and odd datasets) to unity-valued gridded visibilities. Note that this is only the thermal noise contribution, and the signal-to-error plot shows all of the high ratio detections expected in the foreground-dominated regions.
The signal-to-error shows behaviour consistent with thermal noise away from the contaminated wedge and bandpass harmonics regions.

\begin{figure}[t]
\centering
\subfigure[Cross power spectra with the foreground model included in the estimator.]{
\includegraphics[width=.5\textwidth]{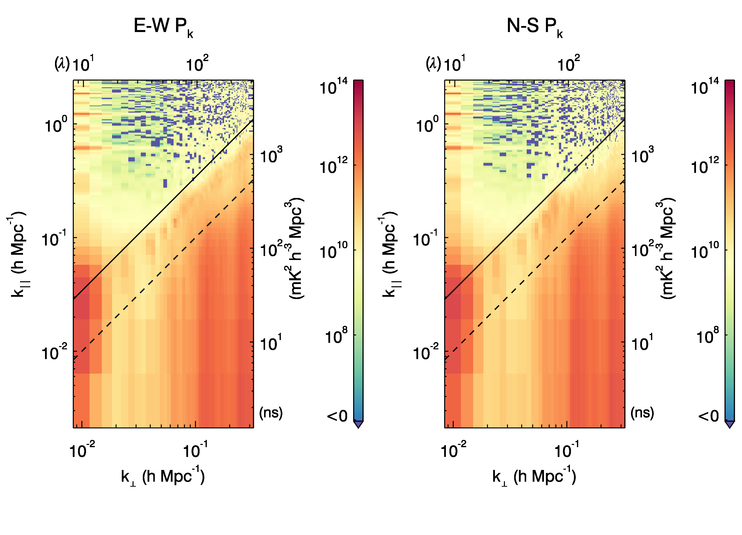}}\\
\subfigure[Errors, when the foregrounds are included in the data covariance matrix.]{
\includegraphics[width=.5\textwidth]{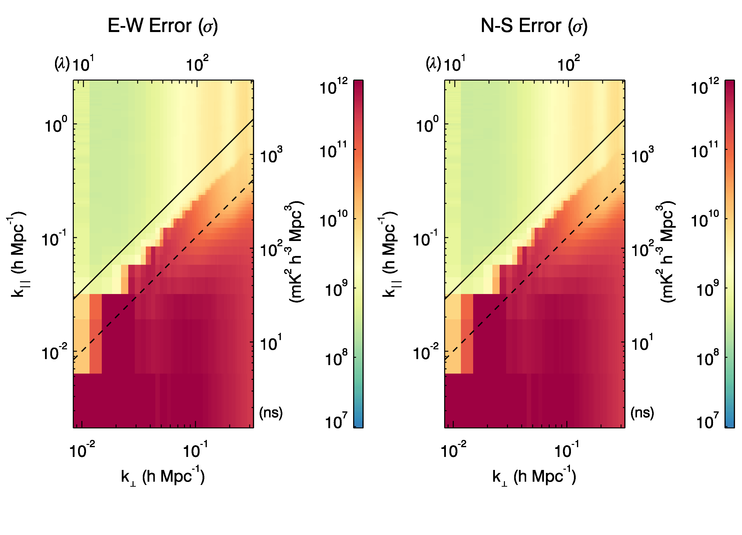}}\\
\subfigure[Signal-to-error ratio plot demonstrating the large suppression of modes within the wedge. Compare with Figure \ref{fig:2d_snr} where no foregrounds are included in the estimator.]{
\includegraphics[width=.5\textwidth]{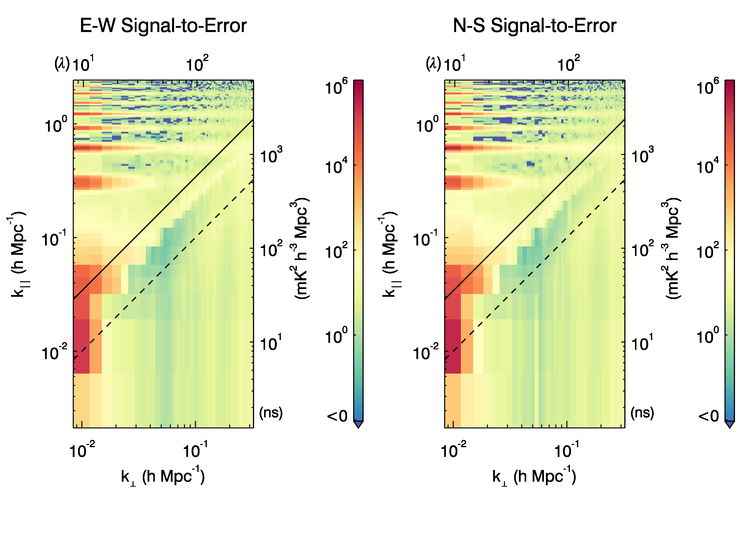}
}
\caption{Cross power spectra, errors and signal-to-error ratios for the two instrumental polarizations, when the foreground model is included in the estimator. The small signal-to-error ratio in the wedge indicates that these modes are highly contaminated and should be down-weighted in the final binning from 2D to 1D.}
\label{fig:2d_power_inv}
\end{figure}

\begin{figure}[t]
\subfigure[$\Delta{z}$=6.2--6.6.]{
\includegraphics[width=.5\textwidth]{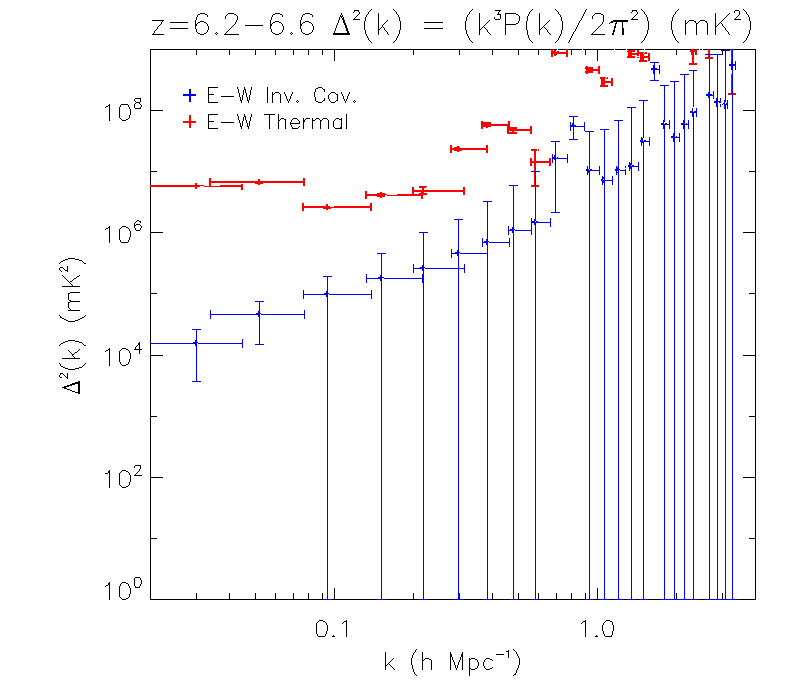}
}
\subfigure[$\Delta{z}$=6.6--7.0.]{
\includegraphics[width=.5\textwidth]{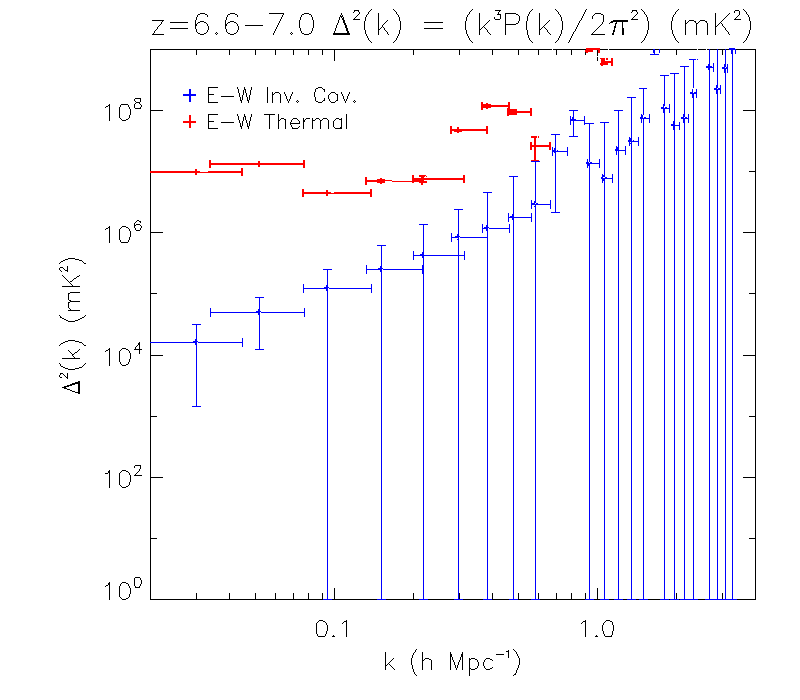}
}\\
\subfigure[$\Delta{z}$=7.0--7.5.]{
\includegraphics[width=.5\textwidth]{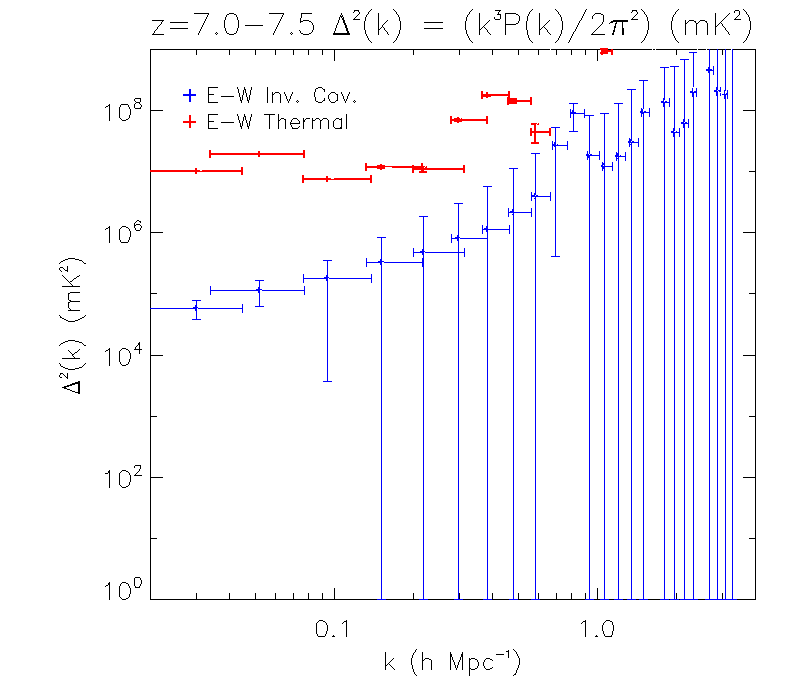}
}
\caption{1D spherically-averaged E-W polarization power spectra for three redshift ranges, corresponding to the upper, mid and lower 10.24~MHz bands of the data. Both a thermal noise-only (red) and a full foreground (blue) data covariance model has been used, and the full parameter space (no wedge excision).}
\label{fig:1d_ps}
\end{figure}

When the foreground contribution is \textit{included} in the data covariance matrix, power is effectively removed from the contaminated regions, and the errors reflect those parts of the parameter space (Figure \ref{fig:2d_power_inv}b). The errors at the locations of the coarse bandpass features are also elevated, but this is not obvious from the plots as shown. The small signal-to-error ratio in the wedge indicates that these modes are highly contaminated and should be down-weighted in the final binning from 2D to 1D. The ratio is close to unity in the wedge, suggesting that the foreground covariance is capturing the contamination appropriately.
These data can then be averaged in cylindrical bins and normalized to obtain the spherically-averaged dimensionless power spectrum. Instead of treating the full bandwidth, which corresponds to a redshift range of 6.2--7.5 and therefore is highly likely to contain signal evolution, we follow \citet{dillon15} and split the coherent data into three contiguous 10.24~MHz bins, corresponding to redshifts $z$=[6.2--6.6], [6.6--7.0], [7.0--7.5], and compute the 1D power spectra for those (Figure \ref{fig:1d_ps}) for the E-W polarization (which shows reduced foreground leakage from the Galaxy). This is effectively using the top and bottom plots from Figure \ref{fig:2d_power_inv} as the power and errors, but with the reduced bandwidth (it is not exactly this process, because the inverse covariance estimator also uses all of the off-diagonal error terms, which are not represented in the 2D error plots). Unlike \citet{dillon15} however, we do not exclude any wedge contribution here, and \textit{instead allow the estimator to `work within the wedge'} and downweight contaminated modes. The only data cut performed is to remove the $k_\bot=0$ and $k_\parallel=0$ bins, which show large contamination and a poor foreground model response.
Both the thermal noise-only and the full thermal noise$+$foreground data covariance matrices are shown. The error bars reflect 95\% confidence regions in both dimensions. Inclusion of the foreground model increases the uncertainties on those modes, corresponding to the substantial contamination in those regions. The full data covariance model increases the uncertainties such that the results are consistent with thermal noise$+$foregrounds across most of parameter space. The `detections' in the $k_\bot=[0.03-0.10]$~h.Mpc$^{-1}$ region are due to power bias from unmodelled foregrounds, recalling that the power bias term from Equation \ref{estimator_equation} is omitted from the estimator.

From these results, we can set 2$\sigma$ upper limits on the EoR power spectrum at the point where the 1D power is at a minimum and we achieve a `detection', corresponding to $k=0.05$h.Mpc$^{-1}$. (We omit the inner detection because the power associated with this bin includes power on the scale of the primary beam size). Table \ref{table:1d_results} lists the upper limits at this wavenumber for the 3 hours of data in the E-W polarization.
\begin{table}[h]
\centering
\begin{tabular}{|l|l|}
\hline 
Redshift & $\Delta_k^2$ (mK$^2$) \\ 
\hline 
$z=[6.2-6.6]$ & 7.6$\times$10$^4$ \\ 
\hline 
$z=[6.6-7.0]$ & 8.8$\times$10$^4$ \\ 
\hline 
$z=[7.0-7.5]$ & 16.5$\times$10$^4$ \\ 
\hline 
\end{tabular} 
\caption{95\% confidence (2$\sigma$) upper limits on the EoR power in three redshift bins at $k=0.05$~h.Mpc$^{-1}$, using the full CHIPS estimator and including all data.}
\label{table:1d_results}
\end{table}
While not competitive with the results from deeper studies with other telescopes, it is consistent with previously published values, with a best detection value of $\Delta_k^2$~=~(275~mK)$^2$ in the redshift range $z=[6.2-6.6]$.

\section{Discussion and conclusions}
The CHIPS estimator is one of several EoR power spectrum estimators being developed and applied within the MWA collaboration, and broadly amongst the community with other low-frequency telescopes. It takes one possible approach to the substantial systematic problems of structured and bright foregrounds, and complex instrumentation. The primary design principles for CHIPS include (1) a full instrumental model (bandpass, frequency- and pointing-dependent primary beam shape, $uv$-sampling, observation-dependent system temperature, instrumental chromaticity); (2) a model-driven foreground component to the data covariance, drawing upon previous observational studies to extract a realistic statistical model for an extragalactic point source population and Galactic synchrotron emission; (3) a maximum-likelihood estimator to tie together the full sky and instrumental information in a consistent and robust framework, and yield a fully-covariant set of output uncertainties.

This approach is not unique, and likely not the full solution to addressing this complex task. In particular, the approach to foregrounds has many options and others have demonstrated the benefits of empirical \citep{dillon15} and blind parametric and non-parametric \citep[e.g.,][]{chapman14} approaches. Ultimately, the approach to foregrounds may require a combination of these techniques. Conversely, an understanding of the full impact of the instrumental and analysis signal chain on the final data product, is crucial for complicated low-frequency telescopes. In this regard, a CHIPS-like approach is likely to be required for most current and future instruments.

The results presented here are from a very small amount of data, and are meant to be interpreted as a proof-of-principle for the general approach. The outputs are visually as we would expect given our understanding of the instrument and expectations of system noise, and quantitatively in one dimension are consistent with previous estimates. In the regions of parameter space where we expect to be thermal noise-limited, this small dataset shows consistency with those expectations.

The foreground model employed here, while multi-component, is still very simple, and includes none of the intricacies of different underlying extragalactic populations (for example, star-forming galaxies and AGN). Being a statistical model, it also lacks the ability to treat any outlier emission that remains after peeling (for example, when peeling sources, the Galactic Centre and main Galactic plane are not treated in this current calibration strategy, but are very bright), leaving additional power in the output power spectrum that would be contained in the unsubtracted bias term. As we probe deeper into the data, we will use the new information gleaned to form better sky models, but this is ultimately an iterative process at low frequencies, where deep observations are sparse.

%\begin{figure}
%\plotone{crosspower_lowband_sept2014.eps}
%\caption{A panel taken from Figure 2 of . 
%See the electronic edition of the Journal for a color version 
%of this figure.\label{fig2}}
%\end{figure}




%\begin{thebibliography}{}

\bibliographystyle{jphysicsB}
\bibliography{pubs.bib}

%\end{thebibliography}

\acknowledgements
\section*{Acknowledgements}
We thank the referee for a very thorough reading of the manuscript, and many excellent suggestions to improve its clarity. This research was supported under Australian Research Council's Discovery Early Career Researcher funding scheme (project number DE140100316). This scientific work makes use of the Murchison Radio-astronomy Observatory, operated by CSIRO. We acknowledge the Wajarri Yamatji people as the traditional owners of the Observatory site. Support for the MWA comes from the U.S. National Science Foundation (grants AST-0457585, PHY-0835713, CAREER-0847753, and AST-0908884), the Australian Research Council (LIEF grants LE0775621 and LE0882938), the U.S. Air Force Office of Scientific Research (grant FA9550-0510247), and the Centre for All-sky Astrophysics (an Australian Research Council Centre of Excellence funded by grant CE110001020). Support is also provided by the Smithsonian Astrophysical Observatory, the MIT School of Science, the Raman Research Institute, the Australian National University, and the Victoria University of Wellington (via grant MED-E1799 from the New Zealand Ministry of Economic Development and an IBM Shared University Research Grant). The Australian Federal government provides additional support via the Commonwealth Scientific and Industrial Research Organisation (CSIRO), National Collaborative Research Infrastructure Strategy, Education Investment Fund, and the Australia India Strategic Research Fund, and Astronomy Australia Limited, under contract to Curtin University. This work was supported by resources provided by the Pawsey Supercomputing Centre with funding from the Australian Government and the Government of Western Australia. This work was supported by resources awarded under Astronomy Australia Ltd’s merit allocation scheme on the gSTAR national facility at Swinburne University of Technology. gSTAR is funded by Swinburne and the Australian Government’s Education Investment Fund. We acknowledge the iVEC Petabyte Data Store, the Initiative in Innovative Computing and the CUDA Center for Excellence sponsored by NVIDIA at Harvard University, and the International Centre for Radio Astronomy Research (ICRAR), a Joint Venture of Curtin University and The University of Western Australia, funded by the Western Australian State government.

\appendix

\section{Expected powers and covariances}
To assess the performance of the estimator, we can take the expected value of the estimate, $\langle \hat{p}_{\alpha} \rangle$, noting that $\langle \vec{x}^T{\bf A}\vec{x} \rangle = {\rm tr}({\bf AC})$,
\begin{eqnarray}
\langle \hat{p}_{\alpha} \rangle &=& \frac{{\rm tr}\left( {\bf C}_{\rm P} {\bf C}^{-1}{\bf C}^{-1} \frac{\partial{\bf C}}{\partial{p_{\alpha}}} \right)}{{\rm tr}\left( {\bf C}^{-1}{\bf C}^{-1}\right)}\\
&=& \frac{\displaystyle\sum_\beta p_{\beta} |({\bf C}^{-1})_{\alpha\beta}|^2}{\displaystyle\sum_\beta |({\bf C}^{-1})_{\alpha\beta}|^2}.
\label{power_estimate}
\end{eqnarray}
The estimate of the power in mode $\alpha$ is therefore a mixture of the cross-power between mode $\alpha$ and others (denoted $\beta$ in the sum), with whitening (decorrelation and weighting) by the data covariance matrix. This expression reduces to this simple form because the derivative,
\begin{equation}
\frac{\partial{\bf C}}{\partial{p_{\alpha}}} = \begin{pmatrix}
	0 & 0 & \cdots & 0\\
	0 & 1 & \cdots & 0\\
	\vdots & \vdots & \ddots & \vdots\\
  	0 & 0 & \cdots & 0
\end{pmatrix}
\end{equation}
has only a single non-zero entry, with a corresponding single non-zero \textit{row} in the quantity, ${\bf C}^{-1} \frac{\partial{\bf C}}{\partial{p_{\alpha}}}$. This is for the case where the cosmological signal is not contaminated, and is confined to a single mode.

The final weighting matrix for the vector of estimates contains all of the weighting and correlation due to the primary beam, and weighting and correlation due to the foreground structure.
\newline
\noindent{\bf Example}

\noindent In the simplest case, there is no foreground or noise contribution, and the covariance matrix, ${\bf C} = {\bf C}_{\rm P}$. Then, equation (\ref{power_estimate}) reduces to:
\begin{eqnarray}
\langle \hat{p}_{\alpha} \rangle &=& \frac{{\rm tr}\left( {\bf C}_{\rm P}^{-1} \frac{\partial{\bf C}}{\partial{p_{\alpha}}} \right)}{{\rm tr}\left( {\bf C}_{\rm P}^{-1}{\bf C}_{\rm P}^{-1}\frac{\partial{\bf C}}{\partial{p_{\alpha}}}\right)}\\
&=& \frac{1/p_{\alpha}}{\displaystyle\sum_\beta |({\bf C}_{\rm P}^{-1})_{\alpha\beta}|^2},
\end{eqnarray}
yielding the true power estimate, weighted by the correlations between mode $\alpha$ and all other modes. In the limit where there are no covariances between modes,
\begin{equation}
\langle \hat{p}_{\alpha} \rangle = \frac{1/p_{\alpha}}{1/{p}_{\alpha}^2} = p_{\alpha},
\end{equation}
yielding an unbiased estimator.

The covariance matrix of the estimator provides a measure of its performance. In the case of a general data covariance, the covariance between powers, $p_{\alpha}$ and $p_{\beta}$ is given by:
\begin{eqnarray}
{\rm cov}(p_{\alpha},p_{\beta}) &\overset{\underset{\mathrm{def}}{}}{=}& \langle \hat{p}_{\alpha}\hat{p}_{\beta} \rangle - \langle \hat{p}_{\alpha} \rangle \langle \hat{p}_{\beta} \rangle\\
&=& \frac{K}{{\rm tr}({\bf E}_\alpha){\rm tr}({\bf E}_\beta)},
\end{eqnarray}
where,
\begin{align}		
K &= 2{\rm tr}({\bf C}^{-1}{\bf C}_\alpha{\bf C}^{-1}{\bf C}_\beta) + {\rm tr}({\bf C}^{-1}{\bf C}_\alpha{\bf C}^{-1}{\bf C}_\beta{\bf C}^{-1}{\bf C}_{N,FG}) \\\nonumber &+ {\rm tr}({\bf C}^{-1}{\bf C}_\alpha{\bf C}^{-1}{\bf C}_{N,FG}{\bf C}^{-1}{\bf C}_\beta) + {\rm tr}({\bf C}^{-1}{\bf C}_\alpha{\bf C}^{-1}{\bf C}_{N,FG}{\bf C}^{-1}{\bf C}_\beta{\bf C}^{-1}{\bf C}_{N,FG}),
\end{align}
with, ${\bf C}_{N,FG} \overset{\underset{\mathrm{def}}{}}{=} {\bf C}_{FG}+{\bf C}_{N}$, and,
\begin{eqnarray}
{\bf E}_\alpha &=& {\bf C}^{-1}{\bf C}^{-1} \frac{\partial{\bf C}}{\partial{p_{\alpha}}}\\
{\bf C}_\alpha &=& \frac{\partial{\bf C}}{\partial{p_{\alpha}}}.
\end{eqnarray}
We have used the expression for the covariance of a zero-mean bilinear quadratic form,
\begin{equation}
{\rm cov}(\vec{V}_1^\dagger {\bf A}_1 \vec{V}_2,\vec{V}_1^\dagger {\bf A}_2 \vec{V}_2) = 2{\rm tr}({\bf A}_1{\bf C}{\bf A}_2{\bf C}),
\end{equation}
where ${\bf A}_1$ and ${\bf A}_2$ are general matrices, and the datasets have means $\vec{V}_1$ and $\vec{V}_2$ and covariance ${\bf C}$.

Given that we are using derived data as input into the estimator (coherently-averaged visibilities, rather than individual measured visibilities), the squaring operates on a small number of visibilities, yielding a $\chi^2$-distribution for the data. (Although the quadratic form formally sums power over the whole $uv$ range, in practise the localization of the beam makes most added powers have zero weight. This is by design to obtain a mostly-diagonal covariance matrix, and coupling only between neighbouring $uv$ points.) The covariance expression described above is therefore an input to an underlying skewed distribution.

\end{document}